\documentclass[runningheads]{llncs}
\usepackage{ulem}
\usepackage{amssymb,amsmath}
\usepackage{paralist}
\usepackage{multirow}
\usepackage{url}
\usepackage[all]{xy}
\usepackage[ruled,vlined,linesnumbered,noend]{algorithm2e}
\usepackage{booktabs}
\usepackage[dvips]{graphicx}
\usepackage[varg]{txfonts}
\usepackage{color}
\usepackage{textcomp}
\usepackage{cases}

\usepackage{latexsym}
\usepackage[varg]{txfonts}

 \urldef{\mailsa}\path|{alfred.hofmann,
brigitte.apfel, ursula.barth, christine.guenther,|
\urldef{\mailsb}\path|ingrid.haas, frank.holzwarth, anna.kramer,
leonie.kunz, nicole.sator,|
\urldef{\mailsc}\path|erika.siebert-cole, peter.strasser,
lncs}@springer.com|

\begin{document}

\title{Scalable Continual Top-$k$ Keyword Search\\ in Relational Databases}
\author{Yanwei Xu\inst{1}}
\institute{Department of Computer Science and Technology, Tongji
University, Shanghai, China\\
}
\maketitle

\begin{abstract}
\emph{Keyword search in relational databases} has been widely studied in recent years because it does not require users neither to master a certain structured query language nor to know the complex underlying database schemas.
Most of existing methods focus on answering \emph{snapshot keyword queries} in static databases.
In practice, however, databases are updated frequently, and users may have long-term interests on specific topics.
To deal with such a situation, it is necessary to build effective and efficient facility in a database system to support
\emph{continual keyword queries}.

In this paper, we propose an efficient method for answering continual top-$k$ keyword queries over relational databases.
The proposed method is built on an existing scheme of keyword search on relational data streams, but incorporates the ranking mechanisms into the query processing methods and makes two improvements to support efficient top-$k$ keyword search in relational databases.
Compared to the existing methods, our method is more efficient both in computing the top-$k$ results in a static database and in maintaining the top-$k$ results when the database continually being updated.
Experimental results validate the effectiveness and efficiency of the proposed method.

\end{abstract}

\section{Introduction}
\label{s:introduction}

With the proliferation of text data available in relational databases, simple ways to exploring such information effectively are of increasing importance.
\emph{Keyword search in relational databases}, with which a user specifies his/her information need by a set of keywords, is a popular information retrieval method because the user needs to know neither a complex query language nor the underlying database schemas.
It has attracted substantial research effort in recent years, and a number of methods have been developed
\cite{dbxplorer,discover,banks,irstyle,banks2,blinks,sigmod06keyword,progressiveICDE,spark,ease08}.

\begin{figure}[htb]

\centering
\begin{minipage}[t]{120mm}
\begin{center}
{\footnotesize
\begin{tabular}{|c|p{105mm}|}
\multicolumn{2}{l}{\bf Papers}\\\hline
\emph{pid} & \multicolumn{1}{|c|}{\emph{title}}\\\hline
$p_1$& ``Leveraging Identity-Based Cryptography for Node ID Assignment in Structured \uwave{P2P} Systems.'' \\\hline
$p_2$& ``\uwave{P2P} or Not \uwave{P2P}?: In \uwave{P2P} 2003'' \\\hline
$p_3$& ``A System for Predicting Subcellular Localization.'' \\\hline
$p_4$& ``Logical Queries over Views: Decidability.'' \\\hline
$p_5$&  ``A conservative strategy to protect \uwave{P2P} file sharing systems from pollution attacks.'' \\\hline
$\cdots$ & \multicolumn{1}{|c|}{$\cdots$} \\\hline
\end{tabular}
}
\end{center}
\end{minipage}

{\centering
\begin{minipage}[t]{50mm}
{\centering
{\footnotesize
\begin{tabular}{|c|c|}
\multicolumn{2}{l}{\bf Authors}\\\hline
  \emph{aid}& \emph{name}\\
\hline
  $a_1$& ``\uwave{James} Chen'' \\
\hline
  $a_2$& ``Saikat Guha''\\
\hline
  $a_3$& ``\uwave{James} Bassingthwaighte''\\
\hline
  $a_4$& ``Sabu T.''\\
\hline
  $a_5$& ``\uwave{James} S. W. Walkerdines''\\
 \hline
  $\cdots$ & $\cdots$\\
\hline
\end{tabular}
}}
\end{minipage}
\begin{minipage}[t]{50mm}
{\centering
{\footnotesize

\begin{tabular}{|c|c|c|c|c|c|c|c|c|c|}
\multicolumn{10}{l}{\bf Writes}\\\hline
  \emph{wid} &$w_1$ & $w_2$& $w_3$ & $w_4$& $w_5$ &  $w_6$  &$w_7$ & $w_8$ &   $\cdots$  \\
  \hline
  \emph{aid} & $a_1$ & $a_2$ & $a_3$ &  $a_1$ &$a_5$ &  $a_3$ &$a_2$ &  $a_2$& $\cdots$ \\
  \hline
  \emph{pid} &$p_2$ & $p_1$ &$p_3$ & $p_4$&$p_5$ &$p_4$ &$p_2$ &$p_5$ & $\cdots$  \\
  \hline
\end{tabular}

}}
\end{minipage}

\ \\
(a) Database (Matched keywords are underlined)

\begin{minipage}[t]{65mm}
\centering
\includegraphics[width=.90\textwidth]{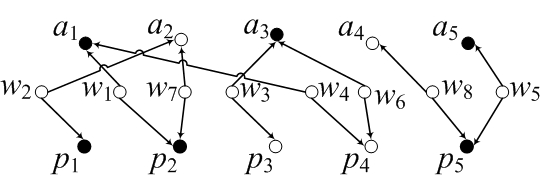}

(b) Tuple connections (Matched tuples\\ are solid circles)
\end{minipage}
\begin{minipage}[t]{45mm}
\centering
\includegraphics[width=.90\textwidth]{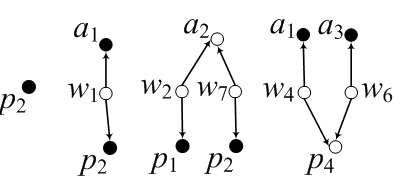}

(c) Examples of query results
\end{minipage}

}

\caption{A sample database with a keyword query ``James P2P''. } \label{fig:runningexample}
\end{figure}

\begin{example}
\label{egKeywordSearch}
Consider a sample publication database shown in Fig.~\ref{fig:runningexample}.
Fig.~\ref{fig:runningexample} (a) shows the three relations \emph{Papers}, \emph{Authors}, and \emph{Writes}.
In the following, we use the initial of each relation name ($P$, $A$, and $W$) as its shorthand.
There are two foreign key references: $\mathit{W}\rightarrow\mathit{A}$ and $\mathit{W} \rightarrow\mathit{P}$.
Fig.~\ref{fig:runningexample} (b) illustrates the tuple connections based on the foreign key references.
For the keyword query ``James P2P'' consisting of two keywords ``James'' and ``P2P'', there are six tuples in the database that contain at least one of the two keywords (underlined in Fig.~\ref{fig:runningexample} (a)).
They can be regraded as the results of the query.
However, they can be joined with other tuples according to the foreign key references to form more meaningful results, several of which are shown in Fig.~\ref{fig:runningexample} (c).
The arrows represent the foreign key references between the corresponding pairs of tuples.
Finding such results which are formed by the tuples containing the keywords is the task of keyword search in relational databases.
As described later, results are often ranked by relevance scores evaluated by a certain ranking strategy. \hfill $\Box$
\end{example}

Most of the existing keyword search methods assume that the databases are static and focus on answering \emph{snapshot} keyword queries.
In practice, however, a database is often updated frequently, and the result of a snapshot query becomes invalid once
the related data in the database is updated.
For the database in Fig.~\ref{fig:runningexample}, if publication data comes continually, new publication records are inserted to the three tables.
Such new records may be more relevant to ``James'' and ``P2P''.
Hence, after getting the initial top-$k$ results, the user may demand the top-$k$ results to reflect the latest database updates.
Such demands are common in real applications.
Suppose a user want to do a top-$k$ keyword search in a Micro-blogging database, which is being updated continually: not only the weblogs and comments are continually being inserted or deleted by bloggers, but also the follow relationship between bloggers are being updated continually.
Thus, a continual evaluation facility for keyword queries is essential in such databases.



For continual keyword query evaluation, when the database is updated, two situations must be considered:
\begin{enumerate}
\item Database updates may change the existing top-$k$ results: some top-$k$ results may be replaced by new ones that are related to the new tuples, and some top-$k$ results may be invalid due to deletions.
\item Database updates may change the relevance scores of existing results because the underlying statistics (e.g., word frequencies) are changed.
\end{enumerate}

In this paper, we describe a system which can efficiently report the top-$k$ results of every monitoring query while the database is being updated continually.
The outline of the system is as follows:
\begin{itemize}
\item
When a continual query is issued, it is evaluated in a pipelined way to find the set of results whose upper bounds of relevance scores are higher than a threshold $\theta$ by calculating the upper bound of the future relevance score for every query result.
\item
When the database is updated, we first update the relevance scores of the computed results, then find the new results whose upper bounds of relevance scores are larger than $\theta$ and delete the results containing the deleted tuples.
\item
The pipelined evaluation of the keyword query is resumed if the number of computed results whose relevance scores are larger than $\theta$ falls below $k$, or is reversed if the above number is much larger than $k$.
\item
At any time, the $k$ computed results whose relevance scores are the largest and are larger than $\theta$ are reported as the top-$k$ results.
\end{itemize}


In Section~\ref{s:preliminaries}, some basic concepts are introduced and the problem is defined.
Section~\ref{s:relwork} discusses related work.
Section~\ref{s:Continual} presents the details of the proposed method.
Section~\ref{s:experiments} gives the experimental results.
Conclusion is drawn in Section~\ref{s:concl}.

\section{Preliminaries}
\label{s:preliminaries}
In this section, we introduce some important concepts for top-$k$ keyword querying evaluation in relational databases.

\subsection{Relational Database Model}
We consider a relational database schema as a directed graph $G_S(V,E)$, called a schema graph, where $V$ represents the set of relation schemas $\{R_1,R_2,\ldots\}$ and $E$ represents the foreign key references between pairs of relation schemas.
Given two relation schemas, $R_i$ and $R_j$, there exists an edge in the schema graph, from $R_j$ to $R_i$, denoted $R_i\leftarrow R_j$, if the primary key of $R_i$ is referenced by the foreign key defined on $R_j$.
For example, the schema graph of the publication database in Fig.~\ref{fig:runningexample} is $\mathit{Papers}\leftarrow\mathit{Write}\rightarrow\mathit{Authors}$.
A relation on relation schema $R_i$ is an instance of $R_i$ (a set of tuples) conforming to the schema, denoted $r(R_i)$.
A tuple can be inserted into a relation.
Below, we use $R_i$ to denote $r(R_i)$ if the context is obvious.

\subsection{Joint-Tuple-Trees (JTTs)}
The results of keyword queries in relational databases are a set of connected trees of tuples, each of which is called a \emph{joint-tuple-tree} ({\emph{JTT}} for short).
A JTT represents how the \emph{matched tuples}, which contain the specified keywords in their text attributes, are interconnected through foreign key references.
Two adjacent tuples of a JTT, $t_i\in r(R_i)$ and $t_j\in r(R_j)$, are interconnected if they can be joined based on a foreign key reference defined on relational schema $R_i$ and $R_j$ in $G_S$ (either $R_i\rightarrow R_j$ or $R_i\leftarrow R_j$).
The foreign key references between tuples in a JTT can be denoted using arrows or notation $\Join$.
For example, the second JTT in Fig.~\ref{fig:runningexample}(c) can be denoted as $a_1\leftarrow w_1\rightarrow p_2$ or $a_1\Join w_1\Join p_2$.
To be a valid result of a keyword query $Q$, each leaf of a JTT is required to contain at least one keyword of $Q$.
In Fig.~\ref{fig:runningexample}(c), tuples $p_1$, $p_2$, $a_1$ and $a_3$ are matched tuples to the keyword query as they contain the
keywords.
Hence, the four JTTs are valid results to the query.
In contrast, $p_1\leftarrow w_2\rightarrow a_2$ is not a valid result because tuple $a_2$ does not contain any required keywords.
The number of tuples in a JTT $T$ is called the \emph{size} of $T$, denoted by $\mathit{size}(T)$.

\subsection{Candidate Networks (CNs)}
Given a keyword query $Q$, the \emph{query tuple set} $R_i^Q$ of relation $R_i$ is defined as $R_i^Q=\{t\in r(R_i)\mid t$ contains some keywords of $Q\}$.
For example, the two query tuple sets in Example~\ref{egKeywordSearch} are $P^Q=\{p_1,p_2,p_5\}$ and $A^Q=\{a_1,a_3,a_5\}$, respectively.
The \emph{free tuple set} $R_i^F$ of a relation $R_i$ with respect to $Q$ is defined as the set of tuples that do not contain any keywords of $Q$.
In Example~\ref{egKeywordSearch}, $P^F=\{p_3,p_4,\ldots\}$, $A^F=\{a_2,a_4,\ldots\}$.
If a relation $R_i$ does not contain text attributes (e.g., relation $W$ in Fig.~\ref{fig:runningexample}), $R_i$ is used to denote $R_i^F$ for any keyword query.
We use $R_i^{QorF}$ to denote a \emph{tuple set}, which may be either $R_i^Q$ or $R_i^F$.

\begin{figure}[ht]
\vspace{-0mm}
\centering
        \includegraphics[width=90mm]{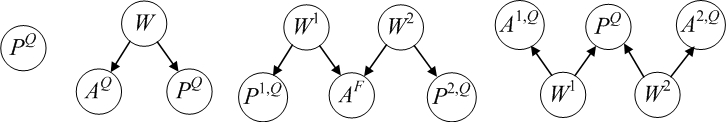}
    \caption{Examples of Candidate Networks}
    \label{figCNExample}
    \vspace{-0mm}
\end{figure}

Each JTT belongs to the result of a relational algebra expression, which is called a \emph{candidate network}~(\emph{CN})~\cite{irstyle,spark,DBIR09}.
A CN is obtained by replacing each tuple in a JTT with the corresponding tuple set that it belongs to.
Hence, a CN corresponds to a join expression on tuple sets that produces JTTs as results, where each join clause $R_i^{QorF}\Join R_{j}^{QorF}$ corresponds to an edge $\langle R_i, R_{j}\rangle$ in the schema graph $G_S$, where $\Join$ represents a equi-join between relations.
For example, the CNs that correspond to two JTTs $p_2$ and $p_2\leftarrow w_1\rightarrow a_1$ in Example~\ref{egKeywordSearch} are $\mathit{P}^Q$ and $\mathit{P}^Q\Join\mathit{W}\Join\mathit{A}^Q$, respectively.
In the following, we also denote $\mathit{P}^Q\Join\mathit{W}\Join\mathit{A}^Q$ as $\mathit{P}^Q\leftarrow\mathit{W}\rightarrow\mathit{A}^Q$.
As the leaf nodes of JTTs must be matched tuples, the leaf nodes of CNs must be query tuple sets.
Due to the existence of $m:n$ relationships (for example, an article may be written by multiple authors), a CN may have multiple occurrences of the same tuple set.
The \emph{size} of CN $C$, denoted as $\mathit{size}(C)$, is defined as the number of tuple sets that it contains.
Obviously, the size of a CN is the same as that of the JTTs it produces.
Fig.~\ref{figCNExample} shows the CNs corresponding to the four JTTs shown in Fig.~\ref{fig:runningexample} (c).
A CN can be easily transformed into an equivalent SQL statement and
executed by an RDBMS.\footnote{ For example, we can transform CN
$\mathit{P}^Q \leftarrow \mathit{W} \rightarrow \mathit{A}^Q$ as: SELECT * FROM W w, P p, A a WHERE w.pid = p.pid AND w.aid = a.aid AND p.pid in ($p_1$, $p_2$, $p_5$) and a.aid in ($a_1$, $a_3$, $a_5$).}

When a continual keyword query $Q = \{w_1, w_2, \ldots, w_l\}$ is specified, the non-empty query tuple set $R_i^Q$ for each relation $R_i$ in the target database are firstly computed using full-text indices.
Then all the non-empty query tuple sets and the database schema are used to generate the set of valid CNs, whose basic idea is to expand each partial CN by adding a $R_i^F$ or $R_i^Q$ at each step ($R_i$ is adjacent to one relation of the partial CN in $G_S$), beginning
from the set of non-empty query tuple sets.
The set of CNs shall be sound/complete and duplicate-free.
There are always a constraint, $CN_{\max}$ (the maximum size of CNs) to avoid generating complicated but less meaningful CNs.
In the implementation, we adopt the state-of-the-art CN generation algorithm proposed in~\cite{SPARKThesis}.

\begin{example}
\label{eg:CNGenerate}
In Example~\ref{egKeywordSearch}, there are two non-empty query tuple sets $\mathit{P}^Q$ and $\mathit{A}^Q$.
Using them and the database schema graph, if $CN_{\max}=5$, the generated CNs are: $CN_1 =
\mathit{P}^Q$, $CN_2 = \mathit{A}^Q$, $CN_3=\mathit{P}^Q\leftarrow \mathit{W}\rightarrow\mathit{A}^Q$, $CN_4=\mathit{P}^{Q}\leftarrow \mathit{W}\rightarrow\mathit{A}^{Q}\leftarrow\mathit{W}\rightarrow\mathit{P}^{Q}$, $CN_5=\mathit{P}^{Q}\leftarrow \mathit{W}\rightarrow\mathit{A}^{F}\leftarrow\mathit{W}\rightarrow\mathit{P}^{Q}$, $CN_6=\mathit{A}^{Q}\leftarrow \mathit{W}^1\rightarrow\mathit{P}^{Q}\leftarrow\mathit{W}\rightarrow\mathit{A}^{Q}$ and $CN_7=\mathit{A}^{Q}\leftarrow \mathit{W}\rightarrow\mathit{P}^{F}\leftarrow\mathit{W}\rightarrow\mathit{A}^{Q}$.
\end{example}

\subsection{Scoring Method}
\label{ss:scoreingMethod}
The problem of \emph{continual top-$k$ keyword search} we study in this paper is to continually report top-$k$ JTTs based on a certain scoring function that will be described below.
We adopt the scoring method employed in \cite{irstyle}, which is an ordinary ranking strategy in the information retrieval area.
The following function $score(T, Q)$ is used to score JTT $T$ for query $Q$, which is based on the TF-IDF weighting scheme:
\begin{equation}
\label{eq:score} {\mathit score}(T,Q)= \frac{\sum_{t\in T}{\mathit tscore}(t,Q)}{{\it size}(T)} ,
\end{equation}
where $t \in T$ is a tuple (a node) contained in $T$. $tscore(t, Q)$ is the \emph{tuple score} of $t$ with regard to $Q$ defined as
follows:
\begin{equation}
\label{eq:tscore} {\mathit tscore}(t,Q) \!=\! \sum_{w\in t\cap Q}
\frac{1+\ln(1+\ln(tf_{t, w}))}{(1-s)+s\cdot \frac{dl_t}{avdl}} \cdot
\ln \left(\frac{N}{df_w+1}\right)
\end{equation}
where $tf_{t, w}$ is the \emph{term frequency} of keyword $W$ in tuple $t$, $df_w$ is the number of tuples in relation $r(t)$ (the relation corresponds to tuple $t$) that contain $W$. $df_w$ is interpreted as the \emph{document frequency} of $W$.
$dl_t$ represents the size of tuple $t$, i.e., the number of letters in $t$, and is interpreted as the \emph{document length} of $t$.
$N$ is the \emph{total number} of tuples in $r(t)$, $avdl$ is the \emph{average tuple size} (\emph{average document length}) in $r(t)$, and $s$~($0 \leq s \leq 1$) is a constant which usually be set to 0.2.

Table~\ref{TupleScores1} shows the tuple scores of the six matched tuples in Example~\ref{egKeywordSearch}.
We suppose all the matched tuples are shown in Fig.~\ref{fig:runningexample}, and the numbers of tuples of the two relations are 150 and 180, respectively.
Therefore, the top-3 results are $T_1=p_2$ ($score=7.04$), $T_2=a_1$ ($score=4.00$) and $T_3=p_2 \leftarrow w_1 \rightarrow a_1$ ($score=3.68$).

\begin{table}[ht]
\caption{Statistics and tuple scores of tuples of $\mathit{P}^Q$ and $\mathit{A}^Q$}
\centering
{\footnotesize
  \begin{tabular}{|l||c|c|c|c|c|c|}
    \hline
    \multicolumn{1}{|c||}{Tuple Set} & \multicolumn{3}{c|}{$\mathit{P}^Q$} &\multicolumn{3}{c|}{$\mathit{A}^Q$}\\
    \hline
    \multirow{2}{*}{Statistics}& $N$&$df_{\mathrm{P2P}}$& $avdl$ & $N$ & $df_{\mathrm{James}}$ & $avdl$ \\
    \cline{2-7}
    & 150&3& 57.8 &170 &3 & 14.6 \\
    \hline
    \vspace{-3mm}
    &\multicolumn{6}{|c|}{}\\
    \hline
    Tuple & $p_1$ & $p_2$ & $p_5$ & $a_1$ & $a_3$ & $a_5$ \\
    \hline
    $dl$ & 88& 28 & 83 & 10 & 22 & 23\\
     \hline
      $tf$ & 1 & 3 & 1 & 1 & 1 & 1\\
    \hline
    $tscore$ & 3.28 & 7.04 & 3.33 & 4.00 & 3.40 & 3.36\\
    \hline
  \end{tabular}
  }
  \label{TupleScores1}
  \vspace{-0mm}
\end{table}

The score function in Eq. (\ref{eq:score}) has the property of \emph{tuple monotonicity}, defined as follows.
For any two JTTs $T=t_1\Join t_2\Join\ldots\Join t_l$ and $T'=t'_1\Join t'_2\Join\ldots\Join t'_l$ generated from the same CN $C$, if for any $1\leq i\leq l$, $\mathit{tscore}(t,Q)\leq \mathit{tscore}(t',Q)$, then we have $\mathit{score}(T,Q)\leq \mathit{score}(T',Q)$.
As shown in the following discussion, this property is relied by the existing top-$k$ query evaluation algorithms.

\section{Related Work}
\label{s:relwork}

\subsection{Keyword Search in Relational Databases}
Given $l$-keyword query $Q=\{w_1,w_2,\cdots,w_l\}$, the task of keyword search in a relational database is to find structural information constructed from tuples in the database~\cite{2010Survey}.
There are two approaches.
The \emph{schema-based approaches} \cite{dbxplorer,discover,irstyle,sigmod06keyword,spark,Skws,KDynamicJournal}
in this area utilize the database schema to generate SQL queries which are evaluated to find the structures for a keyword query.
They process a keyword query in two steps.
They first utilize the database schema to generate a set of relation join templates (i.e., the CNs), which can be interpreted as select-project-join views.
Then, these join templates are evaluated by sending the corresponding SQL statements to the DBMS for finding the query results.
\cite{discover} proved how to generate a complete set of CNs when the $\mathit{CN}_{\max}$ has a user-given value and discussed several query processing strategies when considers the common sub-expressions among the CNs.
\cite{dbxplorer,discover,Skws,KDynamicJournal} all focused on finding all JTTs, whose sizes are $\leq\mathit{CN}_{\max}$, which contain all $l$ keywords, and there is no ranking involved.
In \cite{irstyle} and \cite{spark}, several algorithms are proposed to get top-$k$ JTTs.
We will introduce them in detail in Section~\ref{ss:topK}.

The \emph{graph-based methods} \cite{banks,progressiveICDE,banks2,blinks,ease08,Communities} model and materialize the entire database as a directed graph where the nodes are relational tuples and the directed edges are foreign key references between tuples.
Fig.~\ref{egKeywordSearch}(b) shows such a database graph of the example database.
Then for each keyword query, they find a set of structures (either Steiner trees~\cite{banks}, distinct rooted trees~\cite{banks2}, $r$-radius Steiner graphs~\cite{ease08}, or multi-center subgraphs~\cite{Communities}) from the database graph, which contain all the query keywords and are connected by the paths in database graph.
Such results are found by graph traversals that start from the nodes that contain the keywords.
For the details, please refer the survey papers~\cite{2010Survey,Suvery2010_2}.
The materialized data graph should be updated for any database changes; hence this model is not appropriate to the databases that change frequently~\cite{Suvery2010_2}.
Therefore, this paper adopts the schema-based framework and can be regarded as an extension for dealing with continual keyword search.

\subsection{Top-$k$ Keyword Search in Relational Databases}
\label{ss:topK}

DISCOVER2 \cite{irstyle} proposed the~\emph{Global-Pipelined} (\emph{GP}) algorithm to get the top-$k$ results which are ranked by the IR-style ranking strategy shown in Section \ref{ss:scoreingMethod}.
The aim of the algorithm is to find a proper order of generating JTTs in order to stop early before all the JTTs are generated.
It employs the \emph{priority preemptive}, \emph{round robin} protocol \cite{preemptive} to find results from each query tuple set prefix in a pipelined way, thus each CN can avoid being fully evaluated.

For a keyword query $Q$, given a CN $C$, let the set of query tuple sets of $C$ be $\{R_1^Q,R_2^Q,\ldots,R_m^Q\}$.
Tuples in each $R_i^Q$ are sorted in non-increasing order of their scores computed by Eq.~\ref{eq:tscore}.
Let $R_i^Q.t_j$ be the $j$-th tuple in $R_i^Q$.
In each $R_i^Q$, we use $R_i^Q.cur$ to denote the current tuple such that the tuples before the position of the tuple are all processed, and we use $R_i^Q.cur\leftarrow R_i^Q.cur+1$ to move $R_i^Q.cur$ to the next position.
$q(t_1,t_2,\ldots,t_m)$ (where $t_i$ is a tuple, and $t_i\in R_i^Q$) denotes the parameterized query which checks whether the $m$ tuples can form a valid JTT.
For each tuple $R_i^Q.t_j$, we use $\overline{score}(C.R_i^Q.t_j,Q)$ to denote the upper bound score for all the JTTs of $C$ that contain the tuple $R_i^Q.t_j$, defined as follows:
\begin{equation}
\label{eq:scoreTuple}
\overline{score}(C.R_i^Q.t_j,Q)=\frac{t_j.\mathit{tscore}+\sum_{1\leq i'\leq m\wedge i'\neq i}{C.R_{i'}^Q.t_1.\mathit{tscore}}}{\mathit{size}(C)}
\end{equation}
According to the tuple monotonicity property of Eq.~(\ref{eq:score}) and the sorting order of tuples, among the unprocessed tuples of $C.R_i^Q$, $\overline{score}(C.R_i^Q.{cur},Q)$ has the maximum value.

Algorithm~{GP} initially mark all tuples in $C.R_i^Q$ ($1\leq i\leq m$) of each CN $C$ as un-processed except for the top-most ones.
Then in each while iteration (one round), the un-processed tuple which maximizes the $\overline{score}$ value is selected for processing.
Suppose tuple $C_0.R_s^Q.{cur}$ maximizes $\overline{score}$, processing $C_0.R_s^Q.{cur}$ is done by joining it with the processed tuples in the other query tuple sets of $C_0$ to find valid JTTs:
all the combinations as $(t_1,t_2,\ldots,t_{s-1},R_s^Q.{cur},t_{s+1}\ldots,t_m)$ are tested, where $t_i$ is a processed tuple of $C_0.R_i^Q$ ($1\leq i\leq m$, $i\neq s$).
If the $k$-th relevance score of the found results is larger than $\overline{score}$ values of all the un-processed tuples in all the CNs, it can stop and output the $k$ found results with the largest relevance scores because no results with higher scores can be found in the further evaluation.



One drawback of the GP algorithm is that when a new tuple $C.R_i^Q.{cur}$ is processed, it tries all the combinations of processed tuples $(t_1,t_2,\ldots,t_{s-1},t_{s+1}\ldots,t_m)$ to test whether each combination can be joined with $C.R_i^Q.{cur}$.
This operation is costly due to extremely large number of combinations when the number of processed tuples becomes large \cite{KeywordSearchBook10}.
SPARK \cite{spark} proposes the \emph{Skyline-Sweeping} algorithm to reduce the number of combinations test.
SPARK uses a priority queue $\mathbb{Q}$ to keep the set of seen but not tested combinations ordered by the priority defined as the score of the hypothetical JTT corresponding to each combination.
In each round, the combination in $\mathbb{Q}$ with the maximum priority is tested, then all its adjacent combinations are inserted into $\mathbb{Q}$ but only the combinations that have the high priorities are tested.
SPARK still can not avid testing a huge number of combinations which cannot produce results, though the number of combinations test is highly reduced compared to DISCOVER2.

This paper evaluates the CNs in a pipelined way like \cite{irstyle} and \cite{spark}, but also employs the following two optimization strategies, whose high efficiencies are shown in \cite{discover,Skws,KDynamicJournal}:
\begin{inparaenum}[(1)]
  \item sharing the computational cost among CNs; and
  \item adopting tuple reduction.
\end{inparaenum}

\subsection{Keyword Search in Relational Data Streams}

The most related projects to our paper are \emph{S-KWS}~\cite{Skws} and \emph{KDynamic}~\cite{KDynamic,KDynamicJournal}, which try to find new results or expired results for a given keyword query over an open-ended, high-speed large relational data stream~\cite{2010Survey}.
They adopt the schema-based framework since the database is not static.
This paper deals with a different problem from \emph{S-KWS} and \emph{KDynamic}, though all need to respond to continual queries in a dynamic environment.
\emph{S-KWS} and \emph{KDynamic} focus on finding all query results.
On the contrary, our methods maintain the top-$k$ results, which is less sensitive to the updates of the underlying databases because not every new or expired results change the top-$k$ results.

\emph{S-KWS} maps each CN to a left-deep \emph{operator tree}, where leaf operators (nodes) are tuple sets, and interior operators are joins.
Then the operator trees of all the CNs are compacted into an \emph{operator mesh} by collapsing their common subtrees.
Joins in the operator mesh are evaluated in a bottom-to-top manner.
A join operator has two inputs and is associated with an output buffer which saves its results (partial JTTs).
The output buffer of a join operator becomes input to many other join operators that share the join operator.
A new result that is newly outputted by a join operator will be a new arrival input to those joins sharing it.
The operator mesh has two main shortcomings~\cite{KeywordSearchBook10}:
\begin{inparaenum}[(1)]
  \item only the left part of the operator trees can be shared; and
  \item a large number of intermediate tuples, which are computed by many join operators in the mesh with high processing cost, will not be eventually output in the end.
\end{inparaenum}

\begin{figure}[ht]
\vspace{-5mm}
    \centering
    \begin{minipage}[t]{65mm}
    \centering
    \includegraphics[width=0.90\textwidth,height=26mm]{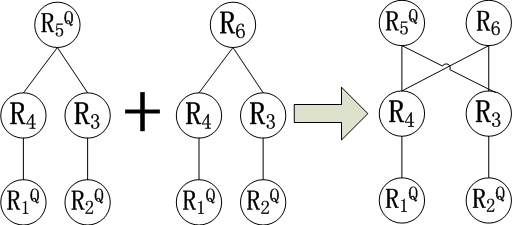}

    (a) $\mathcal{L}$-Lattice of two CNs
    \end{minipage}
    \centering
     \begin{minipage}[t]{26.5mm}
    \centering
    \includegraphics[width=0.95\textwidth]{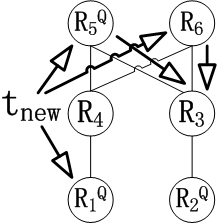}

    (b) Filter phase
    \end{minipage}
    \centering
    \begin{minipage}[t]{26.5mm}
    \centering
    \includegraphics[width=0.95\textwidth]{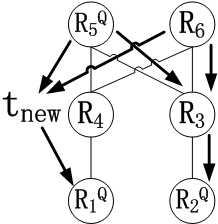}

    (c) Join phase
    \end{minipage}
    \caption{Query processing in \emph{KDynamic}\label{fig:Lattice}}
\end{figure}

For overcoming the above shortcomings of \emph{S-KWS}, \emph{KDynamic} formalizes each CN as a rooted tree, whose root is defined to be the node $r$ such that the maximum path from $r$ to all leaf nodes of the CN is minimized; and then compresses all the rooted trees into a $\mathcal{L}$-Lattice by collapsing the common subtrees.
Fig.~\ref{fig:Lattice}(a) shows the lattice of two hypothetical CNs.
Each node $V$ in the Lattice is also associated with an output buffer, which contains the tuples in $V$ that can join at least one tuple in the output buffer of its each child node.
Thus, each tuple in the output buffer of each top-most node $V$, i.e., the root of a CN, can form JTTs with tuples in the output buffers of its descendants.
The new JTTs involving a new tuple are found in a two-phase approach.
In the filter phase, as illustrated in Fig.~\ref{fig:Lattice}(b), when a new tuple $t_{\rm new}$ is inserted into node $R_4$, \emph{KDynamic} uses selections and semi-joins to check if
\begin{inparaenum}[(1)]
  \item $t_{\rm new}$ can join at least a tuple in the output buffer of each child node of $R_4$; and
  \item $t_{\rm new}$ can join at least a tuple in the output buffers of the ancestors of $R_4$.
\end{inparaenum}
The new tuples that can not pass the checks are pruned; otherwise, in the join phase (shown in Fig.~\ref{fig:Lattice}(c)), a joining process is initiated from each tuple in the output buffer of each root node that can join $t_{\rm new}$, in a top-down manner, to find the JTTs involving $t_{\rm new}$.

In this paper, we incorporate the ranking mechanisms and the pipelined evaluation into the query processing method of \emph{KDynamic} to support efficient top-$k$ keyword search in relational databases.

\section{Continual Top-$k$ Keyword Search in Relational Databases}
\label{s:Continual}

\subsection{Overview}

Database updates bring two orthogonal effects on the current top-$k$ results:
\begin{enumerate}
\item They change the values of $df_w$, $N$, and $avdl$ in Eq.~(\ref{eq:tscore}) and hence change the relevance scores of existing results.
\item New JTTs may be generated due to insertions. Existing top-$k$ results may be expired due to deletions.
\end{enumerate}
Although the second effect is more drastic, the first effect is not negligible for long-term database modifications.
Thus, we can not neglect all the JTTs that are not the current top-$k$ results because some of them have the potential of becoming the top-$k$ results in the future.
This paper solves this problem by bounding the future relevance score of each result.
We use $score^u$ to denote the upper bound of relevance score for each result.
Then, the results whose $score^u$ values are not larger than relevance score of the top-$k$-th results can be safely ignored.

The second challenge is shortage of top-$k$ results because they can be expired due to deletions.
Since the value $k$ is rather small compared to the huge number of all the valid JTTs, the possibility of deleting a top-$k$ result is rather small.
In addition, new top-$k$ results can also be formed by new tuples.
Thus, if the insertion rate is not much smaller than the deletion rate, the possibility of occurring of top-$k$ results shortage would be small.
However, this possibility would be high if the deletion rate is much larger, which can result in frequent top-$k$ results refilling operations.
It worth noting that the top-$k$ results shortage can also be caused by the relevance score changing of results.
Our solution to this problem is to compute the top-$(k+\Delta k)$ ($\Delta k>0$) results instead of the necessary $k$.
$\Delta k$ is a margin value.
Then, we can stand up to $\Delta k$ times of deletion of top results when maintaining the top-$k$ results.
The setting of $\Delta k$ is important.
If $\Delta k$ is too small, it may has a high possibility to refill.
If $\Delta k$ is too large, the efficiency of handling database modifications is decreased.
Instead of analyzing the update behavior of the underlying database to estimate an appropriate $\Delta k$ value, we enlarge $\Delta k$ on each time of top-$k$ results shortage until it reaches a value such that the occurring frequency of top-$k$ results shortage falls below a threshold.

On the contrary, after maintaining the top-$k$ results for a long time, the number of computed top results maybe larger than $(k+\Delta k)$, especially when the insertion rate is high.
In such cases, the top-$k$ results maintaining efficiency is decreased because we need to update the relevance scores for more results and join the new tuples with more tuples than necessary.
As shown in the experimental results, such extra cost is not negligible for long-term database modifications.
Therefore, we need to reverse the pipelined query evaluation if there are too many computed top results.

In brief, when a continual keyword query is registered, we first generate the set of CNs and compact them into a lattice $\mathcal{L}$.
Then, the initial top-$k$ results is found by processing tuples in $\mathcal{L}$ in a pipelined way until the $score^u$ values of the un-seen JTTs are not larger than relevance score of the top-$(k+\Delta k)$-th result (which is denoted by $\mathcal{L}.\theta$).
When maintaining the top-$k$ results, we only find the new results that are with $score^u>\mathcal{L}.\theta$.
The pipelined evaluation of $\mathcal{L}$ is resumed if the number of found results with $score^u>\mathcal{L}.\theta$ falls below $k$, or is reversed if the above number is larger than $(k+\Delta k)$.
The method of computing $score^u$ for results is introduced in Section~\ref{ss:UpperBound}.
Section~\ref{ss:CNEvaluation} and Section~\ref{ss:TopKMaintain} describe our method of computing the initial top-$k$ results and maintaining the top-$k$ results, respectively.
Then, two techniques which can highly improve the query processing efficiency are presented in Section~\ref{ss:CachingJoinedTuples} and Section~\ref{ss:CNClustering}.

\subsection{Computing Upper Bound of Relevance Scores}
\label{ss:UpperBound}

Let us recall the function for computing tuple scores given in Eq.~(\ref{eq:tscore}):
\begin{equation*}
tscore(t, Q) = \sum_{w \in t \cap Q} \frac{1 + \ln(1 + \ln(tf_{t, w}))}{1 - s + s \cdot \frac{dl_t}{avdl}} \cdot \ln \left(\frac{N}{df_w+1}\right) .
\end{equation*}
We assume that the future values of each $\ln\left(\frac{N}{df_{w}+1}\right)$ and $avdl$ both have an upper bound $\ln^{u}\left(\frac{N}{df_w+1}\right)$ and $avdl^{u}$, respectively.
Then, we can derive the upper bound of the future tuple score for each tuple $t$ as:
\begin{equation}
t.\mathit{tscore}^{u}= \sum_{w \in t \cap Q}\frac{1+\ln(1+\ln(tf_{t, w}))}{1-s+s\cdot\frac{dl_t}{avdl^{u}}}\cdot\ln^{u}\left(\frac{N}{df_w+1}\right) .
\end{equation}
Hence, the upper bound of the future relevance score of a JTT $T$ is:
\begin{equation}
\label{scoreUpperBound}
T.\mathit{score}^{u}={\sum_{t\in T}{t.\mathit{tscore}^{u}}}\cdot\frac{1}{\mathit{size}(T)} .
\end{equation}
Note that the function in Eq.~(\ref{scoreUpperBound}) also has the tuple monotonicity property on $\mathit{tscore}^{u}$.

On query registration, each $\ln^{u}\left(\frac{N}{df_w+1}\right)$ is computed as $\ln\left(\frac{N}{df_w(1-\Delta df_w)+1}\right)$, and each $avdl^{u}$ is computed as $avdl(1+\Delta avdl)$, where $\Delta df_w$ and $\Delta avdl$ both are set as small values ($=1\%$).
When maintaining the top-$k$ results, we continually monitor the change of statistics to determine whether all the $\ln\left(\frac{N}{df_w+1}\right)$ and $avdl$ values below their upper bounds.
At each time that any $\ln \left(\frac{N}{df_w+1}\right)$ or $avdl$ value exceeds its upper bound, the $\Delta df_w$ or $\Delta avdl$ is enlarged until the frequencies of exceeding the upper bounds fall below a small number.

\begin{example}
Table~\ref{fig:tscoreBound} shows the $\mathit{tscore}^{u}$ values of the six matched tuples in Example~1 by setting $\Delta df_w=20\%$ and $\Delta avdl=10\%$.
Hence, $T_1.\mathit{score}^{u}=7.42$, $T_2.\mathit{score}^{u}=4.23$ and $T_3.\mathit{score}^{u}=3.88$.
\end{example}

\begin{table}[ht]
\vspace{-9mm}
\caption{Upper bounds of tuple scores}
\centering
\footnotesize{
  \begin{tabular}{|c||c|c|c|c|c|c|}
    \hline
    Tuple & $a_1$  & $a_3$ & $a_5$ & $p_1$ & $p_2$ & $p_5$ \\
 \hline
    $\mathit{tscore}^{u}$ & 4.23 & 3.64 & 3.60 & 3.52 & 7.42 & 3.57\\
    \hline
  \end{tabular}}
\vspace{-9mm}
  \label{fig:tscoreBound}
\end{table}

\subsection{Finding Initial Top-$k$ Results}
\label{ss:CNEvaluation}

\begin{figure}
\vspace{-3mm}
\centering{
    \centering{
    \includegraphics[width=85mm]{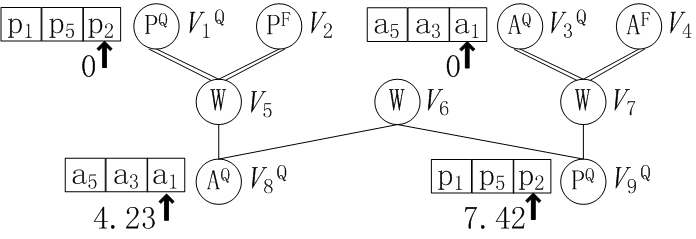}
    }
    }
    \vspace{-2mm}
    \caption{The constructed lattice of the seven CNs in Example~\ref{eg:CNGenerate}}
    \label{fig:LatticeAll}
\vspace{-5mm}
\end{figure}

Fig.~\ref{fig:LatticeAll} shows the $\mathcal{L}$-lattice of the seven CNs in Example~\ref{eg:CNGenerate}.
We use $V_i$ to denote a node in $\mathcal{L}$.
Particularly, $V_i^Q$ denotes a lattice node of query tuple set, and $V_i^Q.R^Q$ denotes the query tuple set of $V_i^Q$.
The dual edges between two nodes, for instance, $V_1^Q$ and $V_5$, indicate that $V_5$ is a dual child of $V_1^Q$.
A node $V_i$ in $\mathcal{L}$ can belongs to multiple CNs.
We use $V_i.\mathit{CN}$ to denote the set of CNs that node $V_i$ belongs to.
For example, $V_8^Q.\mathit{CN}=\left\{CN_2,CN_3,CN_6,CN_7\right\}$.
Tuples in each query tuple set $V_i^Q.R^Q$ are sorted in non-increasing order of $\mathit{tscore}^{u}$.
We use $V_i^Q.cur$ to denote the current tuple such that the tuples before the position of the tuple are all processed, and we use $V_i^Q.cur\leftarrow V_i^Q.cur+1$ to move $V_i^Q.cur$ to the next position.
Initially, for each node $V^Q$ in $\mathcal{L}$, $V_i^Q.cur$ is set as the top tuple in $V_i^Q.R^Q$.
In Fig.~\ref{fig:LatticeAll}, $V_i^Q.cur$ of the four nodes are denoted by arrows.
For a node $V_i$ that is of a free tuple set $R_i^F$, we regard all the tuples of $R_i^F$ as its processed tuples for all the times.
We use $V_i.output$ to indicate the output buffer of $V_i$, which contains its processed tuples that can join at least one tuple in the output buffer of each child node of $V_i$.
Tuples in $V_i.output$ are also referred as the outputted tuples of $V_i$.

In order to find the top-$k$ results in a pipelined way, we need to bound the $\mathit{score}^{u}$ values of the un-found results.
For each tuple $t_j$ of $V_i^Q.R^Q$, the maximal $\mathit{score}^{u}$ values of JTTs that $t_j$ can form is defined as follows:
\begin{equation}
\label{eq:scoreuUpperBound}
\overline{score^{u}}\left(V_i^Q,t_j,Q\right)=
\begin{cases}
0,\quad \text{a child node of $V_i^Q$ has empty output buffer}, \\
\max_{C\in V_i^Q.CN}\left(\overline{score^{u}}\left(C.R^Q.t_j,Q\right)\right),\quad \text{otherwise}
\end{cases}
\end{equation}
where $\overline{score^{u}}\left(C.R^Q.t_j,Q\right)$ indicates the maximal $\mathit{score}^u$ for all the JTTs of $C$ that contain tuple $t_j$, and is obtained by replacing $\mathit{tscore}$ in Eq.~(\ref{eq:scoreTuple}) with $\mathit{tscore}^u$.
If a child of $V_i^Q$ has empty output buffer, processing any tuple at $V_i^Q$ can not produce JTTs; hence $\overline{score^{u}}\left(V_i^Q,t_j,Q\right)=0$ in such cases, which can choke the processing tuples at $V_i^Q$ until all its child nodes have non-empty output buffers.
According to Eq.~(\ref{eq:scoreuUpperBound}) and the tuples sorting order, among the un-processed tuples of $V_i^Q.R^Q$, $\overline{score^{u}}\left(V_i^Q,V_i^Q.{cur},Q\right)$ has the maximum value.
We use $\overline{score^{u}}\left(V_i^Q,Q\right)$ to denote $\overline{score^{u}}\left(V_i^Q,V_i^Q.{cur},Q\right)$.
In Fig.~\ref{fig:LatticeAll}, $\overline{score^{u}}\left(V_i^Q,Q\right)$ values of the four $V_i^Q$ nodes are shown next to the arrows.
For example, $\overline{score^{u}} \left(V_8^Q, Q\right)=\max_{C\in\left\{CN_2,CN_3,CN_6,CN_7\right\}} \left(\overline{score^{u}}\left(C.A^Q.a_1,Q\right)\right)=4.23$.

Algorithm~\ref{alg:EvalStatic} outlines our pipelined algorithm of evaluating the lattice $\mathcal{L}$ to find the initial top-$k$ results, which is similar to the GP algorithm.
Lines~\ref{declareQueue2}-\ref{InitVQCursor} are the initialization step to sort tuples in each query tuple set and to initialize each $V_i^Q.cur$.
Then in each while iteration (lines~\ref{WhileStart2}-\ref{WhileEnd2}), the un-processed tuple in all the $V^Q$ nodes that maximizes $\overline{score^{u}}$ is selected to be processed.
Processing the selected tuples is done by calling the procedure $Insert$.
Algorithm~\ref{alg:EvalStatic} stops when $\max_{V_i^Q\in\mathcal{L}}{\overline{score^{u}}(V_i^Q,Q)}$ is not larger than the relevance score of the top-$(k+\Delta k)$-th found results.
The procedure $Insert(V_i,t)$ is provided in \emph{KDynamic}, which updates the output buffers for $V_i$ (line~\ref{UpdateVOutput}) and all its ancestors (lines~\ref{updateFatherBegin}-\ref{updateFatherEnd}), and finds all the JTTs containing tuple $t$ by calling the procedure $EvalPath$ (line~\ref{findnewJTTsEnd}).
We will explain procedure $Insert$ using examples later.
The recursive procedure $EvalPath(V_i,t,path)$ is provided in \emph{KDynamic} too, which constructs JTTs using the outputted tuples of $V_i$'s descendants that can join $t$.
The stack $path$, which records where the join sequence comes from, is used to reduce the join cost.

\begin{algorithm}
\caption{$\mathit{EvalStatic\text{-}Pipelined}$ (lattice $\mathcal{L}$, the top-$k$ value $k$, $\Delta k$)\label{alg:EvalStatic}}
\SetKwInOut{Input}{input}
\BlankLine
    $\mathit{topk}\leftarrow\emptyset$: the priority queue for storing found JTTs ordered by $score$\label{declareQueue2}\;
    Sort tuples of each $V_i^Q.R^Q$ in non-increasing order of $\mathit{tscore}^{u}$\label{sortTuples}\;
    \lForEach{\normalfont{node $V_i^Q$ in $\mathcal{L}$}}
    {
        let $V_i^Q.{cur}\leftarrow V_i^Q.R^Q.t_1$\label{InitVQCursor}\;
    }
    \While{$\max_{V_i^Q\in\mathcal{L}}{\overline{score^{u}}(V_i^Q,Q)}>topk[k+\Delta k].score$}
    {\label{WhileStart2}
        Suppose $\overline{score^{u}}(V_0^Q,Q)=\max_{V_i^Q\in\mathcal{L}}{\overline{score^{u}}(V_i^Q,Q)}$\;
        $path\leftarrow\emptyset$\tcp*[r]{A stack which records the join sequence}
        $Insert(V_0^Q,V_0^Q.{cur})$\label{callInsert}\tcp*[r]{Processing tuple $V_0^Q.{cur}$ at $V_0^Q$}
        $V_0^Q.{cur}\leftarrow V_0^Q.cur+1$\label{WhileEnd2}\;
    }
    Output the first $k$ results in $\mathit{topk}$\;
    $\mathcal{L}.\theta\leftarrow topk[k+\Delta k].score$\label{setThetaValue}\;

    \BlankLine
    \DontPrintSemicolon
    Procedure $Insert$(lattice node $V_i$, tuple $t$)\;
    \PrintSemicolon
    \If{\normalfont{$t\not\in V_i.output$ and $t$ can join at least one outputted tuple of every child of $V_i$}}
    {\label{checkIftCanOutput}
        Insert $t$ into $V_i.output$\label{UpdateVOutput}\;
    }
    \If{$t\in V_i.output$}
    {
        Push ($V_i$,$t$) to $path$\;
        \lIf{\normalfont{$V_i$ is a root node}}
        {   $\mathit{topk}\leftarrow\mathit{topk}\bigcup EvalPath(V,t,path)$\label{findnewJTTsEnd}\;
        }
        \ForEach{\normalfont{father node of $V_i$, $V_{i'}$ in $\mathcal{L}$}}
        {\label{updateFatherBegin}
            \lForEach{\normalfont{tuple $t'$ belongs to $V_{i'}$ that can join $t$}}
            {\label{selectTuplesCanJoint}
                $Insert(V_{i'},t')$\label{updateFatherEnd}\;
            }
        }
        Pop ($V$,$t$) from $path$\;
    }
\BlankLine
\DontPrintSemicolon
Procedure $EvalPath$(lattice node $V_i$, tuple $t$, stack $path$)\;
\PrintSemicolon
$\mathcal{T}\leftarrow\{t\}$\tcp*[r]{The set of found JTTs}
\ForEach{\normalfont{child node of $V_i$, $V_{i'}$ in $\mathcal{L}$}}
{
    $\mathcal{T'}\leftarrow\emptyset$\tcp*[r]{The set of JTTs that rooted at tuples of node $V_{i'}$}
    \If{$V_{i'}\in path$}
    {
        let $t'$ be the tuple of node $V_{i'}$ that is stored in $path$\;
        $\mathcal{T'}\leftarrow EvalPath(V_{i'},t',path)$\;
    }
    \Else
    {
        \ForEach{\normalfont{tuple $t'\in V_{i'}.output$ that join $t$}}
        {
            $\mathcal{T'}\leftarrow \mathcal{T'}\bigcup EvalPath(V_{i'},t',path)$\tcp*[r]{Union the JTTs that rooted at different tuples of $V_{i'}$}
        }
    }
    $\mathcal{T}\leftarrow \mathcal{T}\times \mathcal{T'}$\tcp*[r]{Compute the Cartesian Product}
}
\Return $\mathcal{T}$\;
\end{algorithm}

\begin{figure}
\vspace{-6mm}
    \centering{
    \includegraphics[width=110mm]{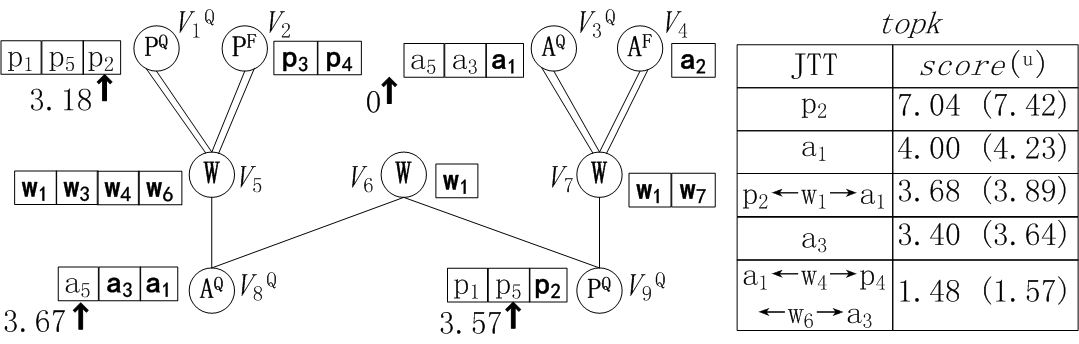}
    }
    \caption{After finding the top-$3$ results (tuples in the output buffers are shown in bold)}
    \label{fig:LatticeAll2}
    \vspace{-5mm}
\end{figure}

\vspace{-0mm}
\begin{example}
In the first round, tuple $V_9^Q.p_2$ is processed by calling $Insert(V_9^Q,p_2)$.
Since $V_9^Q$ is the root node of $CN_1$, $EvalPath$ is called and JTT $T_1=p_2$ is found.
Then, for the two father nodes of $V_9^Q$, $V_6$ and $V_7$, $V_6.output$ is not updated because $V_8^Q.output=\emptyset$, $V_7.output$ is updated to $\{w_1, w_7\}$ because $p_2$ can join $w_1$ and $w_7$.
And then, for the two father nodes of $V_7$, $V_3^Q$ and $V_4$, $V_3^Q.output$ is not updated since $V_3^Q$ has no processed tuples, $V_4.output$ is set as $\{a_2\}$ because there is only one tuple $a_2$ in $A^F$ that can join $w_1$ and $w_7$.
Since $V_4$ is the root node (of $CN_5$), $EvalPath(V_4,a_2,path)$ is called but no results are found because the only one found JTT $p_2\leftarrow w_7\rightarrow a_2\leftarrow w_7\rightarrow p_2$ is not a valid result.
After processing tuple $V_9^Q.p_2$, $\overline{score^{u}} \left(V_3^Q, Q\right)=3.82$ and $\overline{score^{u}} \left(V_9^Q, Q\right)=3.57$.
In the second round, tuple $V_8^Q.a_1$ is processed, which finds results $T_2=a_1$ and $T_3=p_2 \leftarrow w_1 \rightarrow a_1$.
Then, $V_2.output=\{p_4\}$, $V_5.output=\{w_1,w_4\}$, $V_6.output=\{w_1\}$, $\overline{score^{u}} \left(V_1^Q, Q\right)=3.18$, and $\overline{score^{u}} \left(V_8^Q, Q\right)=\overline{score^{u}}\left(CN_3.A^Q.a_3,Q\right)=3.69$.
In the third-fifth rounds, tuples $V_3^Q.a_1$, $V_3^Q.a_3$ and $V_3^Q.a_5$ are processed, which insert $a_1$ into $V_3^Q.output$ and no results found.
In the sixth round, tuple $V_8^Q.a_3$ is processed, which finds results $a_3$ and $a_1\leftarrow w_4\rightarrow p_4\leftarrow w_6\rightarrow a_3$.
Then, Algorithm~\ref{alg:EvalStatic} stops because the relevance score of the third result in the queue $\mathit{topk}$ (suppose $\Delta k=0$) is larger than all the $\overline{score^{u}} \left(V_i^Q, Q\right)$ values.
Fig.~\ref{fig:LatticeAll2} shows the snapshot of $\mathcal{L}$ after finding the top-3 results.
Thus, $\theta=3.68$ after the evaluation.
\end{example}

After the execution of Algorithm~\ref{alg:EvalStatic}, $score^u$ values of all the un-found results are not larger than $\mathcal{L}.\theta$.
Results in the queue $topk$ can be categorized into three kinds.
The first kind are the $(k+\Delta k)$ results that are with $score^u\geq\mathcal{L}.\theta$, which are the initial top-$(k+\Delta k)$ results.
The second kind are with $score<\mathcal{L}.\theta$ and $score^u\geq\mathcal{L}.\theta$, which are called the \emph{potential} top-$(k+\Delta k)$ results because they have the potential to become the top-$(k+\Delta k)$ results.
The third kind are with $score^u\leq \mathcal{L}.\theta$.
As shown in the experiment, the results of the last kind may have a large number.
However, we can not discard them because some of them may become the first two kinds when maintaining the top-$k$ results.

\subsection{Maintaining Top-$k$ Results}
\label{ss:TopKMaintain}


Algorithm~\ref{alg:Maintenance} shows our algorithm of maintaining top-$k$ results.
A database update operator is denoted by $\mathit{OP}(t, R_t)$, which represents a tuple $t$ of relation $R_t$ is inserted (if $\mathit{OP}$ is a insertion) or deleted (if $\mathit{OP}$ is a deletion).
Note that the database updates is modeled as deletions followed by insertions.
For a new arrival $\mathit{OP}(t, R_t)$, Algorithm~\ref{alg:Maintenance} first checks whether the $\ln\left(\frac{N}{df_{w}+1}\right)$ and $avdl$ values of relation $R_t$ exceed their upper bounds.
If some $\ln\left(\frac{N}{df_{w}+1}\right)$(s) or $avdl$ exceeds their upper bounds, we enlarge\footnote{The methods of enlarging $\Delta df_w$, $\Delta avdl$ and $\Delta k$ are introduced in detail in the experiments.} the corresponding $\Delta df_w$(s) or $\Delta avdl$ (line~\ref{EnlargeUpperBound}), and then update the $score$ and $\mathit{score}^u$ values for all the tuples in $R_t^Q$ and all the results in the queue $topk$ using the enlarged $\ln\left(\frac{N}{df_{w}+1}\right)$(s) or $avdl$ (line~\ref{updateAllRelevanceScore});
otherwise, we update the relevance scores for the results in $topk$ that are with $score^u\geq\mathcal{L}.\theta$ (line~\ref{updatePortionRelevanceScore}).
Then, we insert $t$ into $\mathcal{L}$ to find the new results if $\mathit{OP}$ is an insertion (lines~\ref{InsertionBegin}-\ref{InsertionEnd}), or delete the expired JTTs and $t$ from $\mathcal{L}$ if $\mathit{OP}$ is a deletion (lines~\ref{deletionBegin}-\ref{deletionEnd}).
Lines~\ref{InsertionBegin}-\ref{deletionEnd} are explained in detail latter.
And then, the $\overline{score^{u}}(V_i^Q,Q)$ of some nodes may be large than $\mathcal{L}.\theta$, which can be caused by three reasons:
\begin{inparaenum}[\upshape(\itshape 1\upshape)]
  \item the upper bound scores of tuples of relation $R_t$ are increased;
  \item the $\overline{score^{u}}(V_i^Q,Q)$ of some nodes are increased from 0 after inserting the new tuple into $\mathcal{L}$; and
  \item new CNs are added into $\mathcal{L}$.
\end{inparaenum}
Therefore, in lines~\ref{PostStart}-\ref{PostEvaluateOneNode}, we process tuples using procedure $Insert$ until all the $\overline{score^{u}}(V_i^Q,Q)$ values are not larger than $\mathcal{L}.\theta$.

\begin{algorithm}[ht]
\caption{$\mathit{Maintain}$(the evaluated lattice $\mathcal{L}$, the top-$k$ value $k$, $\Delta k$)\label{alg:Maintenance}}
\SetKw{Continue}{continue}
\BlankLine
\While{\normalfont{a new database modification $\mathit{OP} (t, R_t)$ arrives}}
    {\label{newDMArrives2}
        \If{\normalfont{Some $\ln\left(\frac{N}{df_{w}+1}\right)$ (or $avdl$) exceed their upper bounds after applying $\mathit{OP}$}}
        {
            Enlarge the corresponding $\Delta df_w$ (or $\Delta avdl$) value(s)\label{EnlargeUpperBound}\;
            Update relevance scores for tuples in $R_t^Q$ and results in $\mathit{topk}$\label{updateAllRelevanceScore}\label{updateTupleScore}\;
        }
        \Else
        {
            Update $score$ for results in $\mathit{topk}$ that are with $score^u\geq\mathcal{L}.\theta$\label{updatePortionRelevanceScore}\;
        }
        \If(\label{InsertionBegin}\tcp*[f]{Insert $t$ into $\mathcal{L}$}){\normalfont{$\mathit{OP}$ is an insertion}}
        {
            \If{\normalfont{$t$ is an un-matched tuple}}
            {
                \lForEach{\normalfont{node $V_i$ in $\mathcal{L}$ that of $R_t^F$ } }
                {
                    $Insert(V_i,t)$\label{InsertNewTupleToFreeTupleSet}\label{insertUnmatchTuple}\;
                }
            }
            \Else
            {
                \lIf{\normalfont{$R_t^Q$ is new}}
                {\label{newNonEmptyTupleSet}
                    add the new CNs into $\mathcal{L}$\;
                }
                Insert $t$ into $R^Q$ in descending order of $\mathit{tscore}^u$\label{insertNewTupleToTupleSet}\;
                \lForEach{\normalfont{$V_i^Q$ that of $R_t^Q$ and has $\overline{score^u}\left(V_i^Q,t,Q\right)>\mathcal{L}.\theta$ } }
                {\label{measureNewTuplePotential}
                    $Insert(V_i^Q,t)$\label{InsertionEnd}\;
                }
            }
        }
        \ElseIf(\label{deletionBegin}\tcp*[f]{Delete $t$ from $\mathcal{L}$}){\normalfont{$\mathit{OP}$ is a deletion}}
        {
            Delete the results that contain $t$ and are with $score^u\geq\mathcal{L}.\theta$ from $topk$\label{deleteResults}\;
            \lIf{\normalfont{$t$ is a matched tuple}}
            {
                remove $t$ from $R_t^Q$\label{deleteTFromTupleSet}\;
            }
            \lForEach{\normalfont{node $V_i$ in $\mathcal{L}$ such that $t\in V_i.output$}}
            {   \label{deleteTupleBegin}
                $Delete(V_i,t)$\label{deletionEnd}\;
            }
        }
         \While{$\max_{V_i^Q\in\mathcal{L}}{\overline{score^{u}}(V_i^Q,Q)}>\mathcal{L}.\theta$}
        {\label{PostStart}
            \lForEach{\normalfont{node $V_i^Q$ that is with $\overline{score^{u}}(V_i^Q,Q)>\mathcal{L}.\theta$}}
            {$Insert(V_i^Q,V_i^Q.{cur})$\label{PostEvaluateOneNode}\;}

        }
        \If(\tcp*[f]{Resume the evaluation of $\mathcal{L}$}){ $|\{T|T\in\mathit{topk}, T.score\geq\mathcal{L}.\theta\}|<k$}
        {\label{Resume}
            Enlarge $\Delta k$ and then resume the execution of $\mathit{EvalStatic\text{-}Pipelined}$\label{EnlargeDeltaK}\;
        }
        \ElseIf{$|\{T|T\in\mathit{topk}, T.score\geq\mathcal{L}.\theta\}|>(k+\Delta k)$}
        {
            $RollBack(\mathcal{L},k,\Delta k)$\label{RollBackEvaluation}\label{rollBack}\tcp*[r]{Reverse the evaluation of $\mathcal{L}$}
        }
        Report the new first $k$ results in $\mathit{topk}$ if they are changed\;
     }

     \BlankLine
    \DontPrintSemicolon
    Procedure $Delete(V_i,t)$\;
    \PrintSemicolon
    Delete $t$ from $V_i.output$\;

    \ForEach{\normalfont{father node of $V_i$, $V_{i'}$ in $\mathcal{L}$}}
    {\label{deleteFatherTupleBegin}
        \ForEach{\normalfont{tuple $t'$ in $V_{i'}.output$ that can join $t$ only}}
        {\label{selectFatherTuplesToDelete}
            $Delete(V_{i'},t')$\label{deleteTupleEnd}\label{deleteFatherTupleEnd}\tcp*[r]{Call $Delete$ recursively}
        }
    }
\end{algorithm}

Finally, in lines~\ref{Resume}-\ref{rollBack}, we count the number of results that are with $score^u\geq\mathcal{L}.\theta$.
If the number is smaller than $k$, $\Delta k$ is enlarged, and then the $\mathit{EvalStatic\text{-}Pipelined}$ algorithm (without the initialization step) is called to further evaluate $\mathcal{L}$.
If the number is larger than $k+\Delta k$, the algorithm $RollBack$, which is described at the end of this subsection, is called to rollback the evaluation of $\mathcal{L}$.
In any case, at the end of handling the $\mathit{OP}$, we have $\max_{V_i^Q\in\mathcal{L}}{\overline{score^{u}}(V_i^Q.t_{cur},Q)}\leq topk[k].score$.
Therefore, the $k$ results in $topk$ that have the largest relevance scores are the top-$k$ results.
We do not process the results in $\mathit{topk}$ that are with $score^u\leq\mathcal{L}.\theta$ in line~\ref{updatePortionRelevanceScore} and line~\ref{deleteResults}, because they can have a large number and do not have the potential to become top-$k$ results.
However, after the execution of lines~\ref{updateAllRelevanceScore} and \ref{EnlargeDeltaK}, $score^u$ of some of them may become larger than $\mathcal{L}.\theta$, because their $score^u$ values may be enlarged in line~\ref{updateAllRelevanceScore} and the $\mathcal{L}.\theta$ may be decreased in line~\ref{EnlargeDeltaK}.
Therefore, all the results in $\mathit{topk}$ need to be considered in lines~\ref{updateAllRelevanceScore} and \ref{EnlargeDeltaK}.
Note that we have to firstly check whether some of them have expired due to deletions.

\begin{figure}
    \vspace{-5mm}
    \centering{
    \includegraphics[width=80mm]{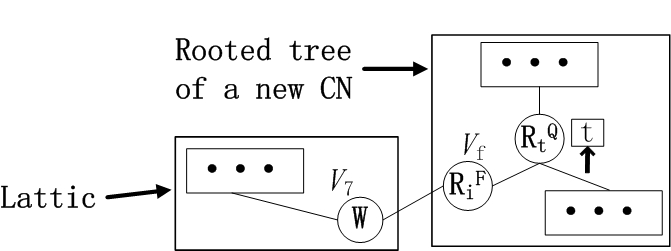}
    }
    \caption{Inserting a new CN into the lattice}
    \vspace{-2mm}
    \label{fig:LatticeInsertNewCN}
    \vspace{-3mm}
\end{figure}

In lines~\ref{InsertionBegin}-\ref{InsertionEnd}, the new tuple $t$ is processed differently according to whether it contains the keywords.
If $t$ is an un-matched tuple, it is inserted into each node of $R_t^F$ using the procedure $Insert$ (line~\ref{InsertNewTupleToFreeTupleSet}).
If $t$ is a matched tuple, inserting it into $\mathcal{L}$ is more complicated.
First, if $t$ introduces a new non-empty query tuple set $R_t^Q$, we add the new CNs involving $R_t^Q$ into the lattice.
Fig.~\ref{fig:LatticeInsertNewCN} illustrates the process of inserting a new CN into the lattice shown in Fig.~\ref{fig:LatticeAll2}.
Assuming that $W\text{---}P^Q$ is the largest common subtree of the new CN and $\mathcal{L}$, and $V_f$ is the father node of $W\text{---}P^Q$ in the new CN, then the new CN is added by setting $V_7$ as the child of $V_f$.
If $V_f$ is a free tuple set and it does not have other child nodes as shown in Fig.~\ref{fig:LatticeInsertNewCN}, $Insert(V_f,t')$ is called for each tuple $t'$ of $V_f$ that can join tuples in $V_7.output$.
Further evaluation at the nodes of the new CN, if necessary, will be done in lines~\ref{PostStart}-\ref{PostEvaluateOneNode}.
Second, $t$ is added into the query tuple set $R^Q$ (line~\ref{insertNewTupleToTupleSet}), and then for each node $V_i^Q$ of $R_t^Q$, $Insert(V_i^Q,t)$ is called when $\overline{score^{u}}\left(V_i^Q.R^Q.t, Q\right)>\mathcal{L}.\theta$ (line~\ref{measureNewTuplePotential}), i.e., $t$ has the potential to form JTTs that are with $score^u>\mathcal{L}.\theta$.

If $OP$ is a deletion, for each node $V_i$ in $\mathcal{L}$ such that $t\in V_i.output$, we delete $t$ from $V_i.output$ using the procedure $Delete$, which is provided by \emph{KDynamic}.
Procedure $Delete$ first removes $t$ from $V_i.output$, and then checks whether some outputted tuples of the ancestors of $V_i$ need to be removed (lines~\ref{deleteFatherTupleBegin}-\ref{deleteFatherTupleEnd}).
For instance, if the tuple $a_3$ is deleted from the lattice node $V_8^Q$ shown in Fig.~\ref{fig:LatticeAll2}, tuples $w_3$ and $w_6$ are deleted from $V_5.output$ too because they can join $a_3$ only, among tuples in $V_8^Q.output$.

Algorithm~\ref{alg:rollBack} outlines out algorithm to reverse the execution of the pipelined evaluation of the lattice.
In the beginning, $\mathcal{L}.\theta$ is set as the relevance score of the $(k+\Delta k)$-th result in the queue $\mathit{topk}$ (line~\ref{ResetThetaValue}).
Then, the processing on each processed tuple $t\in R_i^Q$ that is of $\overline{score^{u}}\left(V_i^Q.R^Q.t,Q\right)\leq\mathcal{L}.\theta$ is reversed (lines~\ref{rollBackBegin}-\ref{rollBackEnd}).
We use $V_i^Q.{cur}-1$ to denote the tuple just before $V_i^Q.{cur}$.
If $t\in R_i^Q.output$, the results involving by $t$ are firstly deleted from $\mathit{topk}$, and then $t$ is deleted from $V_i^Q.output$ by calling the procedure $Delete$.

\begin{algorithm}
\caption{$\mathit{RollBack}$(a lattice $\mathcal{L}$, the top-$k$ value $k$, $\Delta k$)\label{alg:rollBack}}
\BlankLine
$\mathcal{L}.\theta\leftarrow topk[k+\Delta k].score$\label{ResetThetaValue}\;

\ForEach{\normalfont{node $V_i^Q$ in $\mathcal{L}$}}
{
    \While{$\overline{score^{u}}\left(V_i^Q.{cur}-1, Q\right)\leq\mathcal{L}.\theta$}
    {
        \If{$V_i^Q.{cur}-1\in V_i^Q.output$}
        {\label{rollBackBegin}
            Remove the results that are of CNs in $V_i^Q.\mathit{CN}$ and contain tuple $V_i^Q.{cur}-1$ from $topk$\;
            $Delete(V_i^Q,V_i^Q.{cur}-1)$\label{rollBackEnd}\tcp*[r]{Delete from the output buffer}
        }
        $V_i^Q.{cur}\leftarrow V_i^Q.{cur}-1$\;
    }
}
\end{algorithm}

\subsection{Caching Joined Tuples}
\label{ss:CachingJoinedTuples}

In Algorithm~\ref{alg:EvalStatic} and Algorithm~\ref{alg:Maintenance}, procedure $Insert$ and $Delete$ may be called by multiple times upon multiple nodes for the the same tuple.
The core of the two procedures are the \emph{select operations} (or semi-joins \cite{KDynamicJournal}).
For example, in line~\ref{checkIftCanOutput} and line~\ref{selectTuplesCanJoint} of procedure $Insert$, we need to select the tuples that can join $t$ from the output buffer of each child node of $V_i$ and the set of processed tuples of each father node of $V_i$, respectively.
Although such select operations can be done efficiently by the DBMS using indexes, the cost of handling $t$ is high due to the large number of database accesses.
For example, in our experiments, for a new tuple $t$, the maximal number of database accesses can be up to several hundred.


These select operations done for the same tuple $t$ can be done efficiently by sharing the computational cost among them.
Assume a new tuple $w_0$ is inserted into the lattice shown in Fig.~\ref{fig:LatticeAll2}, then procedure $Insert$ is called by three times ($Insert(V_5,w_0)$, $Insert(V_7,w_0)$ and $Insert(V_6,w_0)$) and at most eight selections are done.
All the eight select operations can be expressed using following two relational algebra expressions:
$\pi_{aid}(\sigma_{wid=w_0}(W)\Join\sigma_{aid\in \mathcal{A}_i}(A))$
and $\pi_{pid}(\sigma_{w=w_0}(W)\Join \sigma_{p\in\mathcal{P}_j}(P))$,
where $\mathcal{A}_i$ and $\mathcal{P}_j$ represent the set of tuples in the output buffer of a node or the set of processed tuples of a node.
Since $\mathcal{A}_i$ and $\mathcal{P}_j$ can be different from each other, the eight select operations need to be evaluated individually.
However, if we rewrite the above expressions as
$\pi_{aid}(\sigma_{aid\in \mathcal{A}_i}(\sigma_{wid=w_0}(W)\Join(A)))$ and $\pi_{pid}(\sigma_{p\in\mathcal{P}_j}(\sigma_{w=w_0}(W)\Join(P)))$, the eight select operations would have two common sub-operations: $\sigma_{w=w_0}(W)\Join(A))$ and $\sigma_{w=w_0}(W)\Join(P))$.
If the results of the two common sub-operations can be shared and do selections $\sigma_{aid\in \mathcal{A}_i}$ and $\sigma_{pid\in \mathcal{P}_j}$ in the main memory, the eight select operations can be evaluated involving only two database accesses.

\begin{algorithm}
\caption{$\mathit{CanJoinOneOutputTuple}$(lattice node $V_i$, tuple $t$)\label{alg:canJoinOneOutputTuple}}
\BlankLine
Let $R_i$ be the relation corresponding to the tuple set of $V_i$\;
\If{\normalfont{the tuples of relation $R_i$ that can join $t$ have not been stored}}
{\label{ifHaveStoreJoinedTuples}
    Query the tuples of relation $R_i$ that can join $t$ and store them\label{queryJoinedTuples};
}

\ForEach{\normalfont{tuple $t'$ of the stored tuples of relation $R_i$ that can join $t$}}
{
    \lIf{\normalfont{can find $t'$ in $V_i.output$}}
    {
        \Return \textbf{true}\;
    }
}
\Return \textbf{false}\;
\end{algorithm}

Algorithm~\ref{alg:canJoinOneOutputTuple} shows our procedure to check whether tuple $t$ can join at least one tuple in the output buffer of a lattice node $V_i$, which is called in line~\ref{checkIftCanOutput} of procedure $Insert$.
In line~\ref{queryJoinedTuples}, all the tuples in relation $R_i$ that can join $t$ are queried and cached in the main memory.
This set of cached joined tuples can be reused every time when they are queried.
The procedures for the select operations in line~\ref{selectTuplesCanJoint} of $Insert$ and line~\ref{selectFatherTuplesToDelete} of $Delete$ are also designed in this pattern, which are omitted due to the space limitation.
Note that when the two procedures $Insert$ and $Delete$ are called recursively, select operations done in the above lines are also evaluated by these procedures.
Therefore, for each tuple $t$, a tree of tuples, which is rooted at $t$ and consist of all the tuples than can join $t$, is created.
The tree of tuples can be seen as the cached localization information of $t$.
It is created on-the-fly, i.e., along with the execution of procedures $Insert$ and $Delete$, and its depth is determined by the recursion depth of the two procedures.
The maximum recursion depth of procedures $Insert$ and $Delete$ is $\frac{CN_{\max}}{2}+1$ \cite{KDynamicJournal}, where $CN_{\max}$ indicates the maximum size of the generated CNs.
Hence, the height of this tree of tuples is bounded by $\frac{CN_{\max}}{2}+1$ too.

\begin{figure}
\vspace{-5mm}
    \centering{
    \includegraphics[width=90mm]{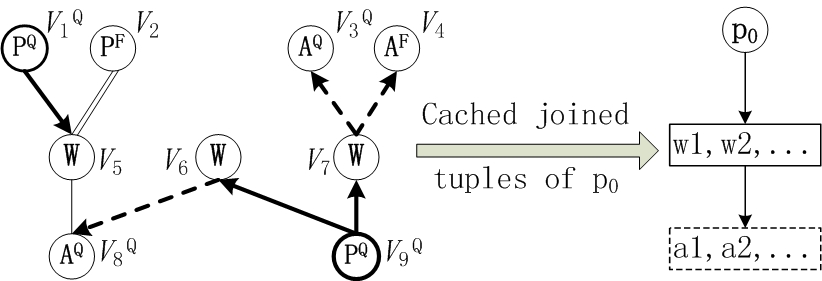}
    }
    \caption{Selections done in $Insert$ and the cached joined tuples for a tuple $p_0$ of $P^Q$}
    \label{fig:LatticeCache}
    \vspace{-4mm}
\end{figure}

Suppose a new tuple $p_0$ of $P^Q$ is inserted into the two nodes of $P^Q$ in the lattice shown in Fig.~\ref{fig:LatticeAll2}, Fig.~\ref{fig:LatticeCache} illustrates the select operations done in the procedure $Insert$ (denoted as arrows in the left part) and the cached joined tuples of $p_0$ (shown in the right part).
For instance, the arrows form $V_9^Q$ to $V_7$ selects the tuples in relation $W$ that can join $p_0$.
The three select operations are denoted by dashed arrows because they would not be done if results of the two select operations, from $V_9^Q$ to $V_7$ and from $V_9^Q$ to $V_6$, are empty.
For the same reason, the stored tuples of relation $A$ that can join $p_0$ are denoted using dashed rectangles.

When computing the initial top-$k$ results, the database is static; hence the cached joined tuples of each tuple unchange and can be reused before the database is updated.
When maintain the top-$k$ results, although the database is continually updated, we can assume the database unchange before $t$ is handled.
However, the cached joined tuples of $t$ is expired after $t$ is handled by Algorithm~\ref{alg:Maintenance}.
As shown in the experimental results, caching the joined tuples can highly improve the efficiency of computing the initial top-$k$ results and maintaining the top-$k$ results.

\subsection{Candidate Network Clustering}
\label{ss:CNClustering}

According to Eq.~(\ref{eq:scoreTuple}), $\overline{score^u}$ values of tuples in different CNs have great differences.
For example, $\overline{score^u}$ values of tuples in $CN_5$ and $CN_7$ are smaller than that of tuples in $CN_3$ due to the large CN size.
In algorithm GP, no tuples or only a small portion are joined in the CNs whose tuples have small $\overline{score}$ values.
If the CNs in Example~\ref{eg:CNGenerate} are evaluated by algorithm GP, $A^Q$ of $\mathit{CN}_7$ and $P^Q$ of $\mathit{CN}_5$ would have no processed tuples.
However, in the lattice, a node $R_i^Q$ can be shared by multiple CNs.
Thus, when inserting a tuple $t$ into $R_i^Q$, $t$ is processed in all the CNs in $R_i^Q.CN$.
As shown in Fig.~\ref{fig:LatticeAll2}, since $V_8^Q$ is shared by $\mathit{CN}_2$, $\mathit{CN}_3$, $\mathit{CN}_6$ and $\mathit{CN}_7$ in the lattice, tuples $a_1$ and $a_3$ are processed in all these four CNs when processing them at $V_8^Q$, which results in \emph{un-needed} operations at nodes $V_2$ and $V_5$ two un-needed results $a_3$ and $a_1\leftarrow w_4\rightarrow p_4\leftarrow w_6\rightarrow a_3$.
We call the operations at $V_2$ and $V_5$ and the two JTTs as un-needed because they wound not occur or be found if the CNs are evaluated separately.
These un-needed operations can cause further un-needed operations when maintaining the top-$k$ results.
For example, we have to join a new unmatched tuple of relation $P$ with four tuples in $V_5.output$.

\begin{figure}
\vspace{-2mm}
    \centering{
    \includegraphics[width=110mm]{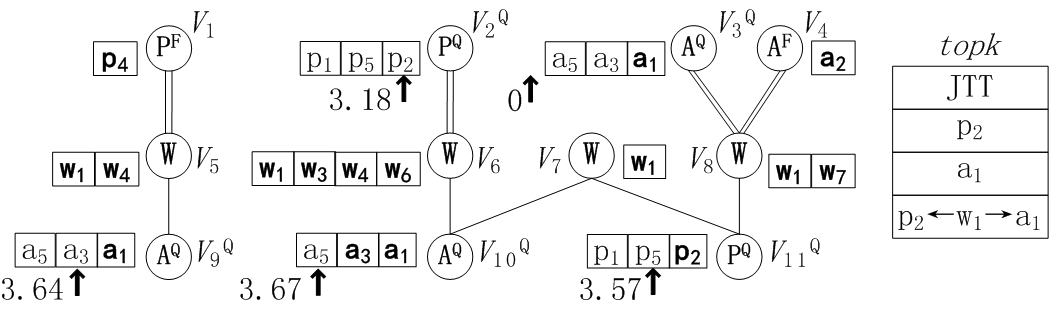}
    }
    \vspace{-3mm}
    \caption{After finding the top-$3$ results if the CNs are clustered into two clusters}
    \label{fig:LatticeAll3}
    \vspace{-3mm}
\end{figure}

The essence of the above problem is that CNs have different potentials in producing top-$k$ results, and then the same tuple set can have different numbers of processed tuples in different CNs if they are evaluated separately.
In order to avoid finding the un-needed results, the optimal method is merely to share the tuple sets that have the same number of processed tuples among CNs when they are evaluated separately.
However, we cannot get these numbers without evaluating the CNs.
As an alternative, we attempt to estimate this number for the tuple sets of each CN $C$ according to following heuristic rules:
\begin{itemize}
  \item If $Max(C)=\displaystyle\frac{\sum_{1\leq i\leq m}C.R_i^Q.t_1.\mathit{tscore}^u}{\mathit{size}(C)}$, which indicates the maximum $score^u$ of JTTs that $C$ can produce, is high, tuple sets of $C$ have more processed tuples.
  \item If two CNs have the same $Max(C)$ values, tuple sets of the CN with larger size have more processed tuples.
\end{itemize}
Therefore, we can cluster the CNs using their $Max(C)\cdot\ln(\mathit{size}(C))$ values, where $\ln(\mathit{size}(C))$ is used to normalize the effect of CN sizes.
Then, when constructing the lattice, only the subtrees of CNs in the same cluster can be collapsed.
For example, $Max(C)\cdot\ln(size(C))$ values of the seven CNs of Example~\ref{eg:CNGenerate} are:
5.15, 2.93, 5.39, 6.84, 5.32, 5.70 and 3.03;
hence they can be clustered into two clusters: $\{CN_2,CN_7\}$ and $\{CN_1,CN_3,CN_4,CN_5,CN_6\}$.
Fig.~\ref{fig:LatticeAll3} shows the lattice after finding the top-3 results if the CNs are clustered, where the three un-needed JTTs in Fig.~\ref{fig:LatticeAll2} can be avoided.
As shown in the experimental section, clustering the CNs can highly improve the efficiency in computing the initial top-$k$ results and handling the database updates.

We cluster the CNs using the $K$-mean clustering algorithm \cite{kmeans}, which needs an input parameter to indicate the number of expected clusters.
We use $\mathit{Kmean}$ to indicate the ratio of this input parameter to the number of CNs.
The value of $\mathit{Kmean}$ represents the trade-off between sharing the computation cost among CNs and considering their different potentials in producing top-$k$ results.
When $\mathit{Kmean}=0$, the CNs is not clustered, then the CNs share the computation cost at the maximum extent.
When $\mathit{Kmean}=1$, all the CNs are evaluated separately.
In our experiments, we find that $\mathit{Kmean}=0.6$ is optimal both for computing the initial top-$k$ results and handling the database updates.


\section{Experimental Study}
\label{s:experiments}
\begin{figure}[ht]
\vspace{-1mm}
\centering{
 \begin{minipage}[t]{100mm}
    \centering{
    \includegraphics[width=0.9\textwidth]{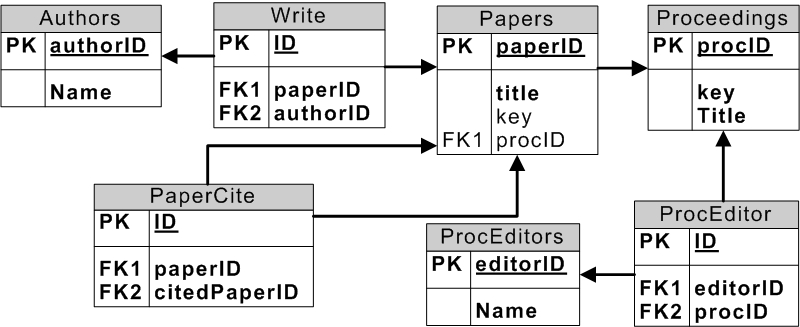}
    }
    \end{minipage}
    }
    \caption{The DBLP schema (PK stands for primary key, FK for foreign key)}
    \vspace{-2mm}
    \label{figdblp}
    \vspace{-3mm}
\end{figure}

We conducted extensive experiments to test the efficiency of our methods.
We use the DBLP dataset\footnote{http://dblp.mpi-inf.mpg.de/dblp-mirror/index.php/}.
Note that DBLP is not continuously growing and is updated on a monthly basis.
The reason we use DBLP to simulate a continuously growing relational dataset is because there is no real growing relational datasets in public, and many studies \cite{irstyle,spark} on top-$k$ keyword queries over relational databases use DBLP.
The downloaded XML file is decomposed into relations according to the schema shown in Fig.~\ref{figdblp}.
The two arrows from \emph{PaperCite} to \emph{Papers} denote the foreign-key-references from \emph{paperID} to \emph{paperID} and \emph{citedPaperID} to \emph{paperID}, respectively.
The DBMS used is MySQL (v5.1.44) with the default ``Dedicated MySQL Server Machine'' configuration.
All the relations use the MyISAM storage engine.
Indexes are built for all primary key and foreign key attributes, and full-text indexes are built for all text attributes.
All the algorithms are implemented in C++.
We conducted all the experiments on a 2.53 GHz CPU and 4 GB memory PC running Windows 7.

\subsection{Parameters}

We use the following five parameters in the experiments:
\begin{inparaenum}[\upshape(\itshape 1\upshape)]
  \item $k$: the top-$k$ value;
  \item $l$: the number of keywords in a query;
  \item $\mathit{IDF}$: the ratio of the number of matched tuples to the number of total tuples, i.e., $\frac{df_w}{N}$;
  \item $\mathit{CN}_{\max}$: the maximum size of the generated CNs; and
  \item $\mathit{Kmean}$: the ratio of the number of clusters of CNs to the number of CNs.
\end{inparaenum}
The parameters with their default values (bold) are shown in Table.~\ref{tbParameters}.
The keywords selected are listed in Table.~\ref{tbKeywords} with their $\mathit{IDF}$ values, where the keywords in bold fonts are keywords popular in author names.
Ten queries are constructed for every $\mathit{IDF}$ value, each of which contains three selected keywords.
For each $l$ value, ten queries are constructed by selecting $l$ keywords from the row of $\mathit{IDF}=0.013$ in Table.~\ref{tbKeywords}.
To avoid generating a small number of CNs for each query, one author name keyword of each $\mathit{IDF}$ value always be selected for each query.


When $k$ grows, the cost of computing the initial top-$k$ results increases since we need to compute more results, and the cost of maintaining the top-$k$ results also increases since there are more tuples in the output buffers of the lattice nodes.
The parameter $\mathit{CN}_{\max}$ has a great impact on keyword query processing because the number of generated CNs increases exponentially while $\mathit{CN}_{\max}$ increases.
And the number of matched tuples increases as $\mathit{IDF}$ and $l$ increase.
Hence, the first four parameters $k$, $l$, $\mathit{IDF}$ and $\mathit{CN}_{\max}$ have effects on the scalability of our method.

\begin{table}[ht]
\vspace{-1mm}
\centering
\begin{minipage}[c]{35mm}
\caption{Parameters\label{tbParameters}}
\centering
{\scriptsize
  \begin{tabular}{|l|p{24mm}|}
    \hline
    Name & Values \\
    \hline
    $k$ & 50, \textbf{100}, 150, 200\\
    \hline
    $l$ & 2, \textbf{3}, 4, 5\\
    \hline
    $\mathit{IDF}$ & 0.003, 0.007, \textbf{0.013}, 0.03\\
    \hline
    $\mathit{CN}_{\max}$ & 4, 5, \textbf{6}, 7\\
    \hline
    $\mathit{Kmean}$ &0, 0.20,0.40,\textbf{0.60},0.80,1\\
    \hline
  \end{tabular}
  }
  \end{minipage}
  \begin{minipage}[c]{80mm}
\caption{Keywords and their $\mathit{IDF}$ values\label{tbKeywords}}
\centering
{\scriptsize
  \begin{tabular}{|p{75mm}|l|}
    \hline
    Keywords & $\mathit{IDF}$ \\
    \hline
    ATM, embedded, navigation, privacy, scalable, Spatial, XML, \textbf{Charles}, \textbf{Eric} & 0.004\\
    \hline
    clustering, fuzzy, genetic, machine, optimal, retrieval, sensor, semantic, video, \textbf{James}, \textbf{Zhang} & 0.007\\
    \hline
    adaptive, architecture, database, evaluation, mobile, oriented, security, simulation, wireless, \textbf{John}, \textbf{Wang} &0.013\\
    \hline
     algorithm, design, information, learning, network, software, time, \textbf{David}, \textbf{Michael} & 0.03\\
    \hline
  \end{tabular}
  }
  \end{minipage}
  \vspace{-4mm}
\end{table}

\subsection{Exp-1: Initial Top-$k$ Results Computation}
In this experiment, we want to study the effects of the five parameters on computing the initial top-$k$ results.
We retrieve the data in the XML file sequentially until number of tuples in the relations reach the numbers shown in Table. \ref{TupleNumbers}.
Then we run the algorithm $\mathit{EvalStatic\text{-}Pipelined}$ on different values of each parameter while keeping the other four parameters in their default values.
We use two measures to evaluate the effects of the parameters.
The first is $\mathit{\#R}$, the number of found results in the queue $\mathit{topk}$.
The second measure is $\mathit{T}$, the time cost of running the algorithm.
Ten top-$k$ queries are selected for each combinations of parameters, and the average values of the metrics of them are reported in the following.
In this experiment, $\Delta df_w$ ($=1\%$), $\Delta avdl$ ($=1\%$) and $\Delta k$ ($=1$) all have very small values because they will be enlarged adaptively when maintaining the top-$k$ results.

\begin{figure}
  \vspace{-4mm}
  \centering
    \begin{minipage}[t]{55mm}
  \centering
  \includegraphics[width=.97\textwidth,height=40mm]{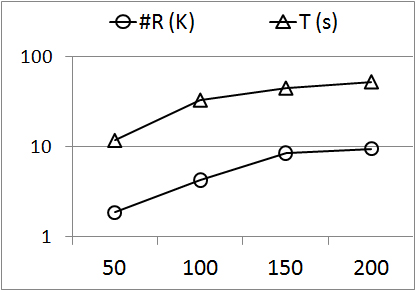}

(a) Varying $k$
  \end{minipage}
  \begin{minipage}[t]{55mm}
  \centering
  \includegraphics[width=.97\textwidth,height=40mm]{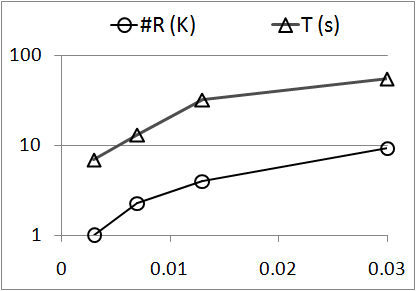}

(b) Varying $IDF$

\ \\

  \end{minipage}

  \begin{minipage}[t]{55mm}
  \centering
  \includegraphics[width=.97\textwidth,height=40mm]{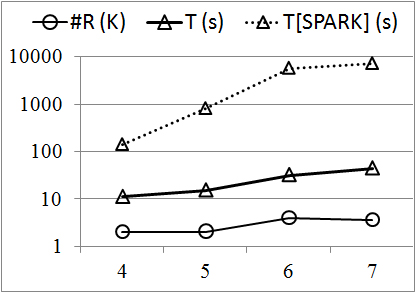}
  (c) Varying $\mathit{CN}_{\max}$ 	

\end{minipage}
  \begin{minipage}[t]{55mm}
  \centering
  \includegraphics[width=.97\textwidth,height=40mm]{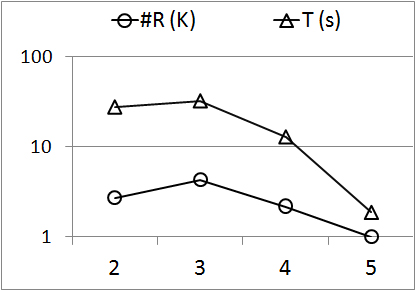}
  (d) Varying $l$

  \ \\

  \end{minipage}

  \begin{minipage}[t]{55mm}
  \centering
  \includegraphics[width=.97\textwidth,height=40mm]{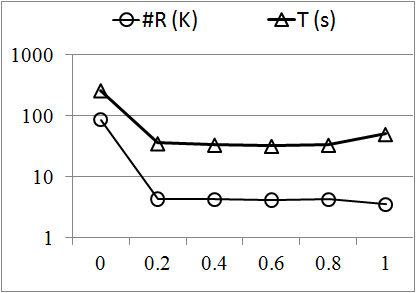}
  (e) Varying $\mathit{Kmean}$

  \ \\

  \end{minipage}
    \begin{minipage}[t]{55mm}
  \centering
  \includegraphics[width=.97\textwidth,height=40mm]{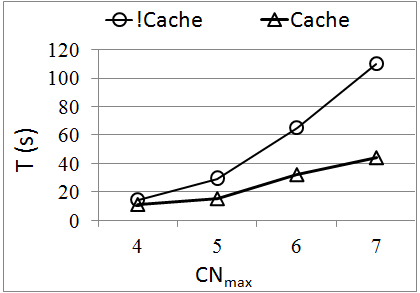}
  (f) Effect of storing joined tuples

  \ \\

  \end{minipage}
  \vspace{-2mm}

  \caption{Experimental results of calculating the initial top-$k$ results}\label{figeffects}
  \vspace{-5mm}
\end{figure}

The main results of this experiment are given in Fig.~\ref{figeffects}.
Note that the units for the $y$-axis are different for the three measures.
Fig.~\ref{figeffects}(a), (b) and (c) show that the two measures all increases as $k$, $idf$ and $CN_{\max}$ grow.
However, they do not show rapid increase in Fig.~\ref{figeffects}(a), (b) and (c), which imply the good scalability of our method.
On the contrary, we can find rapid increase while $CN_{\max}$ grows from the time cost of the method of \cite{spark} in finding the top-$k$ results, which is shown in Fig.~\ref{figeffects}(c) and are denoted by $\mathit{T}[SPARK]$.
Fig.~\ref{figeffects}(c) presents that, compared to the existing method, algorithm $\mathit{EvalStatic\text{-}Pipelined}$ is very efficient in finding the top-$k$ results.
The reason is that evaluating the CNs using the lattice can achieve full reduction because all the tuples in the output buffer of the root nodes can form JTTs and can save the computation cost by sharing the common sub-expressions \cite{KDynamicJournal}.
Fig.~\ref{figeffects}(d) shows that the effect of $l$ seems more complicated: all the two measures may decrease when $l$ increases.
As shown in Fig.~\ref{figeffects}(d), $\mathit{\#R}$ and $T$ even both achieve the minimum values when $l=5$.
This is because the probability that the keywords to co-appear in a tuple and the matched tuples can join is high when the number of keywords is large.
Therefore, there are more JTTs that have high relevance scores, which results in larger $\theta$ and small values of the two measures.

\begin{table}[h]
\vspace{-3mm}
 \caption{Tuple numbers of relations}
 \centering
 {\scriptsize
  \begin{tabular}{|c|c|c|c|c|c|c|}
    \hline
  Papers &	PaperCite & Write &Authors& Proceedings & ProcEditors & ProcEditor \\
\hline
  157,300  	& 9,155 & 400,706 & 190,615  &	2,886 & 1,936 & 1,411 \\
    \hline
  \end{tabular}
  }
  \label{TupleNumbers}
  \vspace{-5mm}
\end{table}

Fig.~\ref{figeffects}(e) presents the changing of the two measures when $\mathit{Kmean}$ varies.
Since the results of the $K$-means clustering may be affected by the starting condition \cite{kmeans}, for each $Kmean$ value, we run Algorithm~\ref{alg:EvalStatic} for 5 times on different starting condition for each keyword query and report the average experimental results.
Note that the algorithm $\mathit{EvalStatic}$ in \emph{KDynamic} corresponding to $\mathit{Kmean}=0$ since there is no CN clustering in \emph{KDynamic}.
From Fig.~\ref{figeffects}(e), we can find that clustering the CNs can highly improve the efficiency of computing the top-$k$ results and the time cost decreases as $\mathit{Kmean}$ increases.
However, when $\mathit{Kmean}=1$, which indicates that all the CNs are evaluated separately, the time cost grows to a higher value than that when $\mathit{Kmean}$ is 0.6 or 0.8.
Therefore, it is important to select a proper $\mathit{Kmean}$ value.
The minimum $T$ in this experiment is achieved on $\mathit{Kmean}=0.6$;
hence the default value of $\mathit{Kmean}$ is 0.6 in our experiments.
As can be seen in the next section, $\mathit{Kmean}=0.6$ also results in the minimum time cost of handling database modifications.

Fig.~\ref{figeffects}(f) compares the time cost of our method in finding the top-$k$ results with that of \emph{KDynamic}, while varying $\mathit{CN}_{\max}$.
The time cost of \emph{KDynamic} is denoted by ``!Cache'' because it does not cache the joined tuples for each tuple.
We can find that caching the joined tuples for each tuple highly improves the efficiency of computing the top-$k$ results.
More important, the improvement increases as $\mathit{CN}_{\max}$ grows.
This is because when $\mathit{CN}_{\max}$ grows, the times of calling the procedure $Insert$ on each tuple increases fast since the number of lattice nodes increases exponentially;
hence the saved cost due to storing the joined tuples of each tuple grows as $\mathit{CN}_{\max}$ grows.

From the curves of $\#R$ in Fig.~\ref{figeffects}, we can find that $\#R$ values are large in all the settings: about several thousand.
Recall that $\mathit{topk}$ contains three kinds of results.
The number of the first kind of results is $k+\Delta k$, which is small compared to the $\#R$ values.
Since $\Delta df_w$ ($=1\%$), $\Delta avdl$ and $\Delta k$ all have very small values, the number of potential top-$(k+\Delta k)$ results in $\mathit{topk}$ is very small ($<10$).
Therefore, the third kind of results, which are with $score^u<\mathcal{L}.\theta$, is in the majority and has a lager number.

\subsection{Exp-2: Top-$k$ Result Maintenance}

In this experiment, we want to study the efficiency of Algorithm~\ref{alg:Maintenance} in maintaining top-$k$ results.
We use the same keyword queries as Exp-1.
After calculating the initial top-$k$ results for them, we sequentially insert additional tuples into the database by retrieving data from the DBLP XML file.
At the same time, we delete randomly selected tuples from the database.
Algorithm~\ref{alg:Maintenance} is used to maintain the top-$k$ results for the queries while the database being updated.
The database update records are read from the database log file;
hence the database updating rate has no directly impact on the efficiency of top-$k$ results maintenance because the database is updated by another process.

\begin{figure}
\vspace{-5mm}
	\centering
  \begin{minipage}[t]{55mm}
  \centering
  \includegraphics[width=0.97\textwidth,height=40mm]{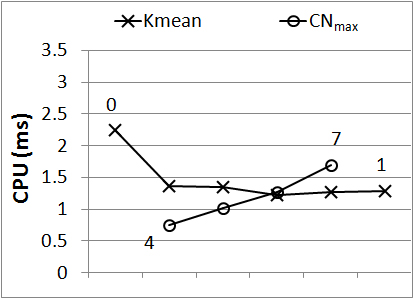}\\
  (a) Time for handling database updates while varying $Kmean$ and $\mathit{CN}_{\max}$

  	\ \\

  \end{minipage}
  \begin{minipage}[t]{55mm}
  \centering
  \includegraphics[width=0.97\textwidth,height=40mm]{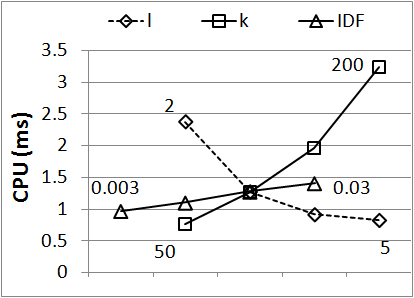}\\
  (b) Time for handling database updates while varying $l$, $k$ and $IDF$
  \end{minipage}

  \centering
  \begin{minipage}[t]{55mm}
  \centering
  \includegraphics[width=0.97\textwidth,height=40mm]{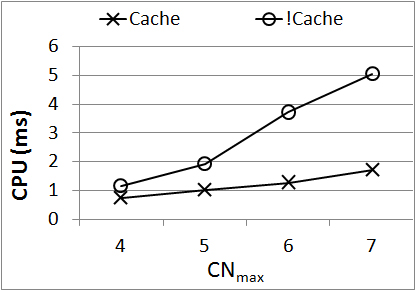}\\
  (c) Effect of storing joined tuples in handling database updates
	\ \\

\end{minipage}
  \centering
  \begin{minipage}[t]{55mm}
  \centering
  \includegraphics[width=0.97\textwidth,height=40mm]{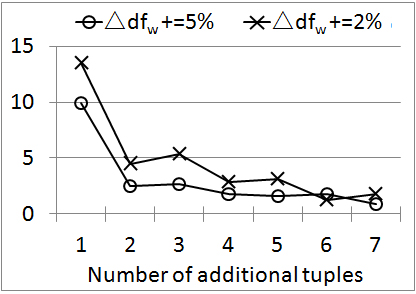}\\
  (d) Changes of the times of enlarging $\Delta df_w$

	\ \\

\end{minipage}

  \centering
  \begin{minipage}[t]{55mm}
  \centering
  \includegraphics[width=0.97\textwidth,height=40mm]{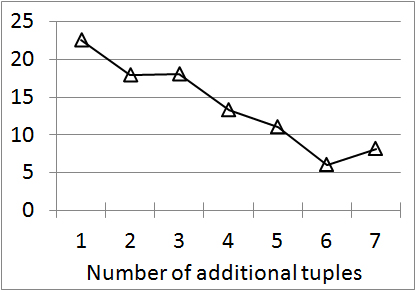}\\
  (e) Changes of the times of calling procedure $RollBack$
	\ \\

\end{minipage}
  \centering
  \begin{minipage}[t]{55mm}
  \centering
  \includegraphics[width=0.97\textwidth,height=40mm]{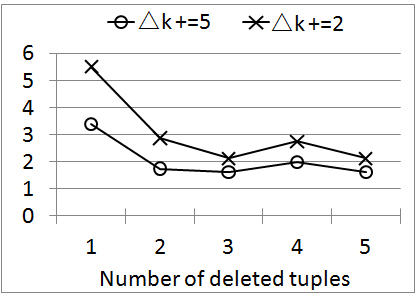}\\
  (f) Changes of the times of enlarging $\Delta k$

	\ \\

\end{minipage}

  \caption{Efficiency of top-$k$ result maintenance}
  \label{figEfficiency}
\vspace{-5mm}
\end{figure}

We first add 713,084 new tuples into the database and delete 250,000 tuples from the database.
The new data is roughly 90 percent of the data used in Exp-1.
The composition of the additional tuples is shown in Table.~\ref{NewTuples}.
Fig.~\ref{figEfficiency}(a) and (b) show the change of the average execution times of Algorithm~\ref{alg:Maintenance} in handling the above database updates when varying the five parameters\footnote{Since it is hard to read in one figure, we split the data of the five parameters into two figures.}, which presents the efficiency of Algorithm~\ref{alg:Maintenance}.
Note that the units for the $x$-axis are different for the five measures, whose minimum and maximum values are labeled in Fig.~\ref{figEfficiency}(a) and (b), and their other values can be found in Table.~\ref{tbParameters}.
We can find that the time cost of handling database updates for the default queries is smaller than 1.5ms.
Comparing Fig.~\ref{figEfficiency}(a) and (b) with the curves of measure $T$ in Fig.~\ref{figeffects} (especially the curves in Fig.~\ref{figeffects}(d) and Fig.~\ref{figeffects}(e)), we can find that the time cost to handle database updates and the time cost to compute the initial top-$k$ results have the same changing trends.
This is because there are more outputted tuples in the lattice when more time is needed to compute the initial top-$k$ results; hence more time is required to do the selections in procedures $Insert$ and $Delete$ and the recursive depthes of them are more larger.
Fig.~\ref{figEfficiency}(c) compares the time cost of our method in handling database updates with that of \emph{KDynamic}, while varying $\mathit{CN}_{\max}$.
The time cost of \emph{KDynamic} is also denoted as ``!Cache''.
We can find that caching the joined tuples for each tuple can also improve the efficiency of handling database updates, and the larger the $\mathit{CN}_{\max}$, the higher the improvement of the efficiency is.

\begin{table}
\vspace{-2mm}
 \caption{Composition of the additional tuples}
 \centering
 {\scriptsize
  \begin{tabular}{|c|c|c|c|c|c|c|}
    \hline
  Papers &	PaperCite & Write &Authors& Proceedings & ProcEditors & ProcEditor \\
\hline
 156,965  	& 20,010 & 411,109 & 111,094  &	3,033 & 3,886 & 6,987 \\
    \hline
  \end{tabular}
  }
  \label{NewTuples}
  \vspace{-2mm}
\end{table}

Secondly, we only insert the 713,084 additional tuples into the database while maintaining top-$100$ results for the default ten keyword queries.
We adopt two different growing rates of $\Delta df_w$: $\Delta df_w+=2\%$ and $\Delta df_w+=5\%$, which mean that when a $\ln\left(\frac{N}{df_w+1}\right)$ exceed its upper bound, the corresponding $\Delta df_w$ value is increased by $2\%$ and $5\%$, respectively.
After inserting each 100,000 additional tuples, we record the average frequency of enlarging $\Delta df_w$ and calling the procedure $RollBack$ for the ten queries, whose changes are shown in Fig.~\ref{figEfficiency}(d) and (e), respectively, whose $x$-axis (with unit of $10^5$) indicate the number of additional tuples.
Note that we do not report the frequency of enlarging $avdl$ because it is very small in the experiment ($<2$).

Fig.~\ref{figEfficiency}(d) shows rapid decrease after inserting the first 100,000 additional tuples.
Although the frequency of enlarging $\Delta df_w$ is larger when the growing rate of $\Delta df_w$ is lower, after inserting 300,000 additional tuples, the times of enlarging $\Delta df_w$, i.e., the times of exceeding the upper bound of $\ln\left(\frac{N}{df_w+1}\right)$, falls below 5 for both the two growing rates of $\Delta df_w$.
After inserting 300,000 additional tuples, the maximum $\Delta df_w$ value of all the relations is 15; hence it is reasonable to set 15 as the maximum value for $\Delta df_w$.
There is only one curve in Fig.~\ref{figEfficiency}(e) because the growing rate of $\Delta df_w$ has no great impact on the times of calling the procedure $RollBack$, which is mainly affected by the frequency of finding new results that are with $score^u>\mathcal{L}.\theta$.
Note that $\mathcal{L}.\theta$ is increased after each time of calling the procedure $RollBack$.
Therefore, the times of calling the procedure $RollBack$ is decreasing since it is more and more harder to find new results that are with $score^u>\mathcal{L}.\theta$.
In order to study the impact of reversing the pipelined evaluation on the efficiency of handling database updates, we also redo the experiment without calling the procedure $RollBack$.
Then, the average time cost of handling database updates is increased by $45.4\%$, which confirms the necessity of reversing the pipelined evaluation.

Then, we delete 500,000 randomly selected tuples from the database after inserting the 713,084 additional tuples.
Two different $\Delta k$ growing rates are adopted: $\Delta k+=2$ and $\Delta k+=5$, which mean that when the number of results that are with $score^u>\mathcal{L}.\theta$ falls below $k$, the corresponding $\Delta k$ value is increased by $2$ and $5$, respectively.
We record the average times of enlarging $\Delta k$ of the ten queries after deleting each 100,000 tuples, whose changes are shown in Fig.~\ref{figEfficiency}(f).
Fig.~\ref{figEfficiency}(f) shows that the frequency of shortage of top-$k$ results falls below a very small number after deleting 200,000 tuples, i.e., after $\Delta k$ being enlarged to about 20.
As indicated by the curve of $k$ in Fig.~\ref{figEfficiency}(b), a large $\Delta k$ value can highly decrease the efficiency of handling database updates.
Therefore, it is reasonable to set the maximum value of $\Delta df_w$ as $20\%$.

\section{Conclusion}
\label{s:concl}

In this paper, we have studied the problem of finding the top-$k$ results in relational databases for a continual keyword query.
We proposed an approach that finds the answers whose upper bounds of future relevance scores are larger than a threshold.
We adopt an existing scheme of finding all the results in a relational database stream, but incorporate the ranking mechanisms in the query processing methods and make two improvements that can facilitate efficient top-$k$ keyword search in relational databases.
The proposed method can efficiently maintain top-$k$ results of a keyword query without re-evaluation.
Therefore, it can be used to solve the problem of answering continual keyword search in databases that are updated frequently.

\vspace{-3mm}
\section*{Acknowledgments}
\vspace{-3mm} This research was partly supported by the National
Natural Science Foundation of China (NSFC) under grant No.~60873040,
863 Program under grant No.~2009AA01Z135. Jihong Guan was also
supported by the ``Shu Guang'' Program of Shanghai Municipal
Education Commission and Shanghai Education Development Foundation.

\vspace{-5mm}

\bibliographystyle{ieicetr}

\end{document}